\documentclass[a4paper,fleqn]{cas-dc}

\usepackage[numbers]{natbib}

\usepackage{placeins}  % Prevents figures from floating beyond sections
\usepackage{graphicx} % <-- Load the graphics package
\usepackage{booktabs}
\usepackage{multirow}
\usepackage{graphicx}
\usepackage{amsmath}
\usepackage{url}
\usepackage{placeins}                 % Prevents figures from floating past sections

\def\tsc#1{\csdef{#1}{\textsc{\lowercase{#1}}\xspace}}
\tsc{EHB}
\tsc{QE}
\tsc{EP}
\tsc{PMS}
\tsc{BEC}
\tsc{DE}
\begin{document}
\let\WriteBookmarks\relax
\def\floatpagepagefraction{1}
\def\textpagefraction{.001}
\shorttitle{Deep learning for artifacts reduction in EEG during simultaneous fMRI}
\shortauthors{K. A. Shahriar et~al.}

\title [mode = title]{Deep Learning for Gradient and BCG Artifacts Removal in EEG During Simultaneous fMRI}                      
%\tnotemark[1,2]

%\tnotetext[1]{This document is the results of the research
%   project funded by the National Science Foundation.}

%\tnotetext[2]{The second title footnote which is a longer text matter
%   to fill through the whole text width and overflow into
%   another line in the footnotes area of the first page.}

\author[1]{K. A. Shahriar}[type=editor,
                        auid=000,bioid=1,
                        %prefix=Dr.,
                        %role=Researcher,
                        orcid=0009-0008-2571-7485]
												
%\cormark[1]
%\ead{kh.ashikshahriar@gmail.com}
%\ead[url]{https://chicago.medicine.uic.edu/profiles/bhuiyan-enamul/}

%\credit{Writing – review \& editing, Writing – original draft, Visualization, Validation, Software, Investigation, Data curation, Statistical analysis, Conceptualization}

\author[2]{E. H. Bhuiyan}[type=editor,
                        auid=000,bioid=1,
                        %prefix=Dr.,
                        %role=Researcher,
                        orcid=0000-0002-3098-1095]
												
\cormark[1]
\ead{bhuiyan@uic.edu}
%\ead[url]{https://chicago.medicine.uic.edu/profiles/bhuiyan-enamul/}

%\credit{Writing – review \& editing, Writing – original draft, Visualization, Validation, Software, Investigation, Data curation, Statistical analysis, Conceptualization}

\author[1,3]{Q. Luo}[type=editor,
                        auid=000,bioid=1,
                        %prefix=Dr.,
                        %role=Researcher,
                        orcid=0000-0002-4169-8447]
												
%\cormark[1]
\ead{qluo@uic.edu}
%\ead[url]{https://chicago.medicine.uic.edu/profiles/luo-qingfei/}

%\credit{Review \& editing, Visualization, Validation, Software}

\author[4]{M. E. H. Chowdhury}[type=editor,
                        auid=000,bioid=1,
                        %prefix=Dr.,
                        %role=Researcher,
                        %orcid=0000-0001-0000-0000]
												]
%\cormark[1]
\ead{mchowdhury@qu.edu.qa}
%\ead[url]{www.jkkrishnan.in}

%\credit{Review \& editing, Validation, AI experts}

\author[1,5,6]{X. J. Zhou}[type=editor,
                        auid=000,bioid=1,
                        %prefix=Dr.,
                        %role=Researcher,
                        orcid=0000000307934925]
												
%\cormark[1]
%\cormark[2]
\ead{xjzhou@uic.edu}
%\ead[url]{https://chicago.medicine.uic.edu/profiles/zhou-xiaohong/}

%\credit{Review \& editing, Visualization, Validation, Project Coordination}
																																											
%\cormark[1]
%\ead{jkk@example.in}
%\ead[url]{www.jkkrishnan.in}

%\credit{Conceptualization of this study, Methodology, Software}

%\address[1]{, Street 129, 1043 NX Amsterdam, The Netherlands}
\affiliation[1]{organization={Department of Electrical and Electronic Engineering},
                addressline={Bangladesh University of Engineering and Technology}, 
                city={Dhaka},
%               citysep={}, % Uncomment if no comma needed between city and postcode
                postcode={1000}, 
                state={Dhaka},
                country={Bangladesh}}

%\address[1]{, Street 129, 1043 NX Amsterdam, The Netherlands}
\affiliation[2]{organization={Center for Magnetic Resonance Research},
                addressline={University of Illinois Chicago}, 
                city={Chicago},
%               citysep={}, % Uncomment if no comma needed between city and postcode
                postcode={60612}, 
                state={IL},
                country={USA}}

\affiliation[3]{organization={Department of Radiology},
                addressline={University of Illinois Chicago}, 
                city={Chicago},
%               citysep={}, % Uncomment if no comma needed between city and postcode
                postcode={60612}, 
                state={IL},
                country={USA}}

\affiliation[4]{organization={Department of Electrical Engineering},
                addressline={Qatar University}, 
                city={Doha},
%               citysep={}, % Uncomment if no comma needed between city and postcode
                postcode={213}, 
%                state={VA},
                country={Qatar}}	

\affiliation[5]{organization={Department of Biomedical Engineering},
                addressline={University of Illinois Chicago}, 
                city={Chicago},
%               citysep={}, % Uncomment if no comma needed between city and postcode
                postcode={60607}, 
                state={IL},
                country={USA}}

\affiliation[6]{organization={Departments of Radiology and Neurosurgery},
                addressline={University of Illinois College of Medicine at Chicago}, 
                city={Chicago},
%               citysep={}, % Uncomment if no comma needed between city and postcode
                postcode={60612}, 
                state={IL},
                country={USA}}	
																																								
%\author[2,4]{Han Thane}[style=chinese]

%\author[2,3]{William {J. Hansen}}[%
%   role=Co-ordinator,
%   suffix=Jr,
%\ead{wjh@example.org}
%\ead[URL]{https://www.university.org}

%\credit{Data curation, Writing - Original draft preparation}

%\affiliation[2]{organization={World Scientific University},
%                addressline={Street 29}, 
%                postcode={1011 NX}, 
%                postcodesep={}, 
%                city={Amsterdam},
%                country={The Netherlands}}
%
%\author[1,3]{T. Rafeeq}
%\cormark[2]
%\fnmark[1,3]
%\ead{t.rafeeq@example.in}
%\ead[URL]{www.campus.in}
%
%\affiliation[3]{organization={University of Intelligent Studies},
%                addressline={Street 15}, 
%               city={Jabaldesh},
%                postcode={825001}, 
%                state={Orissa}, 
%                country={India}}
%
\cortext[cor1]{Enamul Hoque Bhuiyan, Phone: +1-203-214-6948, Fax: +1-312-355-1635, email: bhuiyan@uic.edu}
%\cortext[cor2]{Xiaohong Joe Zhou, Phone: +1-312-413-3979, Fax: +1-312-355-1635, email: xjzhou@uic.edu}
\fntext[fn1]{Xiaohong Joe Zhou (Senior author), Center for Magnetic Resonance Research, University of Illinois Chicago, email: xjzhou@uic.edu.}
%\fntext[fn2]{Another author footnote, this is a very long footnote and
%  it should be a really long footnote. But this footnote is not yet
%  sufficiently long enough to make two lines of footnote text.}

%\nonumnote{This note has no numbers. In this work we demonstrate $a_b$
%  the formation Y\_1 of a new type of polariton on the interface
%  between a cuprous oxide slab and a polystyrene micro-sphere placed
%  on the slab.
%  }

\begin{abstract}
Simultaneous EEG-fMRI recording combines high temporal and spatial resolution for tracking neural activity. However, its usefulness is greatly limited by artifacts from magnetic resonance (MR), especially gradient artifacts (GA) and ballistocardiogram (BCG) artifacts, which interfere with the EEG signal. To address this issue, we used a denoising autoencoder (DAR), a deep learning framework designed to reduce MR-related artifacts in EEG recordings. Using paired data that includes both artifact-contaminated and MR-corrected EEG from the CWL EEG/fMRI dataset, DAR uses a 1D convolutional autoencoder to learn a direct mapping from noisy to clear signal segments. Compared to traditional artifact removal methods like principal component analysis (PCA), independent component analysis (ICA), average artifact subtraction (AAS), and wavelet thresholding, DAR shows better performance. It achieves a root-mean-squared error (RMSE) of 0.0218  0.0152, a structural similarity index (SSIM) of 0.8885 $\pm$ 0.0913, and a signal-to-noise ratio (SNR) gain of 14.63 dB. Statistical analysis with paired t-tests confirms that these improvements are significant (p<0.001; Cohen’s d >1.2). A leave-one-subject-out (LOSO) cross-validation protocol shows that the model generalizes well, yielding an average RMSE of 0.0635 $\pm$ 0.0110 and an SSIM of 0.6658 $\pm$ 0.0880 across unseen subjects. Additionally, saliency-based visualizations demonstrate that DAR highlights areas with dense artifacts, which makes its decisions easier to interpret. Overall, these results position DAR as a potential and understandable solution for real-time EEG artifact removal in simultaneous EEG-fMRI applications.

%\noindent\texttt{\textbackslash begin{abstract}} \dots 
%\texttt{\textbackslash end{abstract}} and
%\verb+\begin{keyword}+ \verb+...+ \verb+\end{keyword}+ 
%which
%contain the abstract and keywords respectively. 

%\noindent Each keyword shall be separated by a \verb+\sep+ command.
\end{abstract}

\begin{graphicalabstract}
\includegraphics[width=1.0\textwidth]{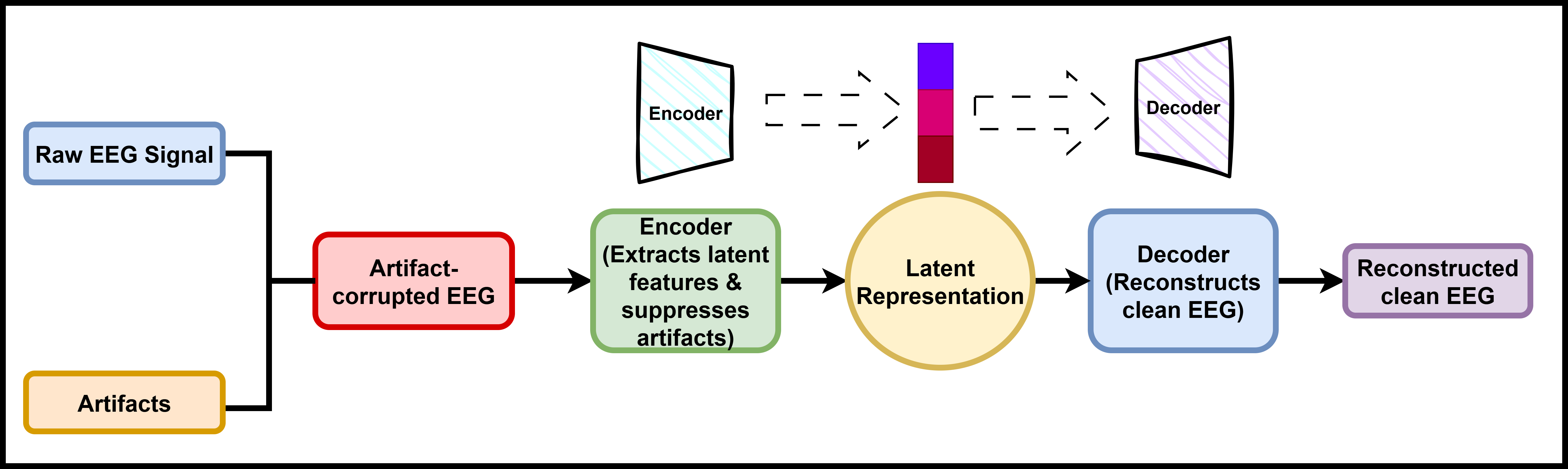}
\end{graphicalabstract}

\begin{highlights}
  \item We deployed DAR, an interpretable 1D convolutional autoencoder for removing artifacts in simultaneous EEG-fMRI recordings.
  \item DAR shows better performance through thorough evaluations compared to traditional and deep learning methods.
  \item We incorporate saliency-based interpretability to build clinical trust and improve scientific understanding.
  \item This study demonstrates robust subject-wise generalization through rigorous cross-validation and ablation testing.
\end{highlights}


\maketitle

\section{Introduction}
\label{sec: Introduction}

Electroencephalography (EEG) is a non-invasive technique for capturing the brain's electrical signals in great detail over time frames accurately, which is valuable for exploring brain wave patterns and cognitive functions, as well as different health issues~\cite{uriguen2015eeg,roy2019deep}. Functional magnetic resonance imaging (MRI), on the other hand, provides detailed spatial information and detects alterations in blood circulation~\cite{freyer2009ultrahigh,logothetis2008we}. The increasing fascination lies in the merging of EEG and fMRI technologies as they enable the capture of electrical and blood flow signals. This integrated approach enhances our comprehension of brain function~\cite{blinowska2021practical,scrivener2021simultaneous,mulert2013simultaneous}.

Recording EEG, within an MRI machine, commonly faces challenges caused by gradient artifacts (GA) and ballistocardiogram interference~\cite{wu2016real, yan2009understanding}. The phenomenon called gradient artifacts arises from the alterations of MRI gradients in imaging processes; on the other hand, motion artifacts related to heart activity lead to what is known as blastocardiography (BCG) artifacts in the magnetic field environment. These unwanted effects can overpower signals by a factor of up to 100 times, making it vital to apply corrections for raw EEG data to be meaningful~\cite{mullinger2013best}.

Traditional methods for removing artifacts, such as average artifact subtraction (AAS)~\cite{gonccalves2007artifact}, optimal basis set (OBS)~\cite{wu2016real}, and adaptive filtering~\cite{jafarifarmand2013artifacts, kher2016adaptive, correa2007artifact}, rely on template-based subtraction. Both independent component analysis (ICA) and principal component analysis (PCA) are commonly employed to deconstruct EEG data into components facilitating manual or semiautomatic removal of artifacts~\cite{chawla2011pca,turnip2014removal,ter2013reduction}. While these techniques can be useful, to some extent, they often need tweaks, rely on linear connections, and might inadvertently weaken actual biological cues.

Lately, in the field of biomedical signals, the cleaning process has been enhanced by the rise of deep learning (DL), which shows potential in capturing intricate and nonlinear patterns effectively~\cite{mcintosh2020ballistocardiogram, yang2018automatic, lin2022ballistocardiogram, duffy2020gradient, mayeli2021automated}. Convolutional neural networks (CNNs), recurrent neural networks (RNNs), and hybrid architectures have been tried for general EEG signal cleaning~\cite{gao2022eeg, chuang2022ic}, but only a few studies explore artifact removal during simultaneous EEG-fMRI, a situation with especially challenging noise issues. Moreover, many DL models lack transparency, which limits how they can be used in clinical and research settings~\cite{samek2017explainable, tjoa2020survey}.

To tackle these issues effectively, we utilized a denoising autoencoder (DAR) to remove artifacts from EEG data. This specific autoencoder is structured with channel 1 dimensional convolutions, aimed at filtering out GA and BCG artifacts from EEG signals captured during fMRI scans. Unlike approaches, the DAR model doesn't depend on predefined artifact patterns or manual identification and removal of components. Instead, it autonomously learns to rectify artifacts from the input data. In order to enhance the interpretability, our approach incorporates a saliency-based analysis, which sheds light on how the network prioritizes areas during artifact elimination.

We evaluated DAR using a publicly available dataset that combines EEG and fMRI data. We compared its effectiveness with methods like ICA and advanced DL models such as U net 2d and ResNet 2d autoencoders in our study. Our assessment involved metrics to measure performance quality visually and statistically through significance testing and exclusion studies on individual subjects to evaluate its applicability in diverse scenarios.

Our research highlights the potential of DAR to improve the quality of EEG artifact removal in studies combining EEG and fMRI simultaneously by reducing noise and enhancing interpretability; DAR shows promise for ensuring reliable EEG results in clinical settings as well. This initial study provides insights into integrating data-driven correction seamlessly into routine simultaneous EEG fMRI scans and ultimately contributes to advancing our comprehension of brain function.

\subsection{Related Work}
\label{subsec: Related Work}

During the process of artifact removal in EEG signal, particularly in cases of simultaneous EEG-fMRI collection, research has investigated both conventional approaches and data-driven methods. While each method yields perspectives on the matter at hand, there are still limitations in place that call for additional enhancements.

Template-based methods, such as AAS and the OBS, are commonly used to correct GA and BCG artifacts~\cite{allen2000method,niazy2005removal}. AAS creates an average artifact template and subtracts it from EEG recordings. Some studies report an up to 90\% reduction in GA amplitude, although residual noise often remains~\cite{freyer2009ultrahigh}. OBS builds on this by using a set of basis functions to account for temporal variations. However, it relies heavily on accurate synchronization and usually achieves only modest improvements in signal-to-noise ratio (SNR) of around 5 to 10 dB~\cite{niazy2005removal}.

Blind source separation (BSS) methods, such as ICA~\cite{srivastava2005ica,turnip2014removal} and PCA~\cite{ter2013reduction}, are still widely used. ICA shows correlation gains of 0.5 to 0.7 between cleaned and reference EEG signals~\cite{srivastava2005ica}, but it requires manual selection of components, which introduces subjectivity and may risk overlooking genuine neural activity~\cite{debener2008properties}. Additionally, ICA performance tends to drop when there is a high overlap of artifacts or non-linear mixing, common in EEG-fMRI settings.

Adaptive filtering techniques use external reference signals, such as ECG or MR gradient triggers, to estimate and remove artifacts dynamically~\cite{correa2007artifact}. These methods can cut artifact power by 60 to 80\%~\cite{roy2019deep}, effectively managing non-stationary interference. However, their success depends heavily on the quality of the reference signal. Poor or inconsistent signals can result in incomplete artifact removal or signal distortion~\cite{kher2016adaptive}.

Recently, DL has emerged as a powerful tool for EEG denoising. It can model complex nonlinear relationships without relying on manually designed features. Models based on CNNs and RNNs have shown strong performance in removing artifacts while preserving important neural dynamics. Hybrid models, like U-Nets and residual networks, have improved results by combining local and global feature representations~\cite{duffy2020gradient,mayeli2021automated,gao2022eeg}.

Despite these advancements, many DL-based approaches tend to focus on isolated artifacts, such as eye blinks or EMG noise, while giving limited attention to GA and BCG artifacts found in EEG-fMRI acquisitions~\cite{duffy2020gradient,mayeli2021automated}. Furthermore, the lack of clarity in many DL models raises concerns, as their black-box nature provides little insight into how they make artifact correction decisions in terms of time or space~\cite{samek2017explainable, tjoa2020survey}.

To deal with these problems effectively, DAR has been developed as a channel-based one convolutional autoencoder tailored to fix EEG fMRI artifacts. Unlike approaches, DAR incorporates interpretability based on saliency, highlighting important regions crucial for noise reduction. This model surpasses deep learning benchmarks in measures such as RMSE, SSIM, and correlation.

\section{Theoretical Background}
\label{sec: Theoretical Background}

The deployment of DAR aims to meet the specific needs of EEG signals and the effects of MRI-induced artifacts. EEG recordings are inherently unstable and show complex, localized changes over time. These signals are also very vulnerable to GA and BCG artifacts, which can be much stronger than the neural activity, sometimes by a factor of 100.

The DAR uses convolutional layers, which are effective at capturing localized features and handling complex nonlinear patterns. Unlike traditional techniques, convolutional layers do not rely on pre-defined artifact templates or outside reference signals. The model features a channel-wise architecture that considers the differences in spatial distribution across EEG electrodes. This allows DAR to develop specific artifact correction mappings for each channel. This is essential due to the varying strength of artifacts and physiological signals across different areas of the scalp.

At the heart of DAR is a convolutional autoencoder framework that learns compressed representations of EEG signals that have been distorted by artifacts. The encoder transforms the noisy input into a simplified latent space that keeps important neural information while reducing noise. The decoder then reconstructs a clean EEG signal from this representation. Unlike approaches based on segmentation or classification, the autoencoder provides a direct mapping from corrupted to clean signals, allowing continuous artifact suppression.

The DAR trains to reduce the difference between the clean reference signal \(x_i\) and the network's output \(f_\Theta(\tilde{x}_i)\) generated from the corrupted input \(\tilde{x}_i\). The objective function is defined as:

\begin{equation}
\mathcal{L}(\Theta) = \frac{1}{N} \sum_{i=1}^{N} \left\| x_i - f_\Theta(\tilde{x}_i) \right\|_1
\end{equation}

Here, \(f_\Theta(\cdot)\) represents the autoencoder mapping defined by \(\Theta\), \(N\) is the number of training segments, and \(\|\cdot\|_1\) shows the L1 norm, which helps achieve strong reconstruction while keeping detailed oscillatory information intact~\cite{bank2023autoencoders}.

The balanced encoder-decoder framework helps maintain both local patterns and the overall shape of the waveforms, which is vital for preserving relevant EEG features. Using L1 loss makes the system less sensitive to large outliers, allowing for the detection of subtle but important neural oscillations.

\begin{figure*}[htbp]
    \centering
    \includegraphics[width=\textwidth]{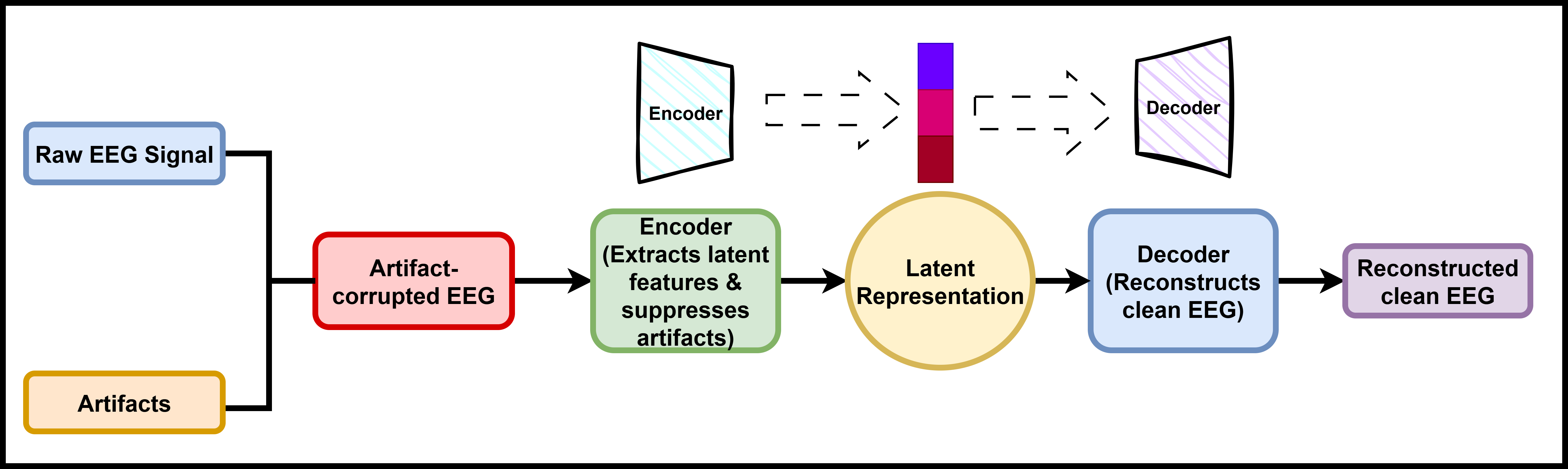}
    \caption{Conceptual illustration of the DAR denoising autoencoder framework. The raw EEG signal is corrupted by MRI-induced artifacts, resulting in an artifact-corrupted EEG segment. The encoder extracts latent features and reduces artifacts, leading to a compact latent representation. The decoder then reconstructs the artifact-free EEG signal from this representation, guided by an L1 reconstruction loss against clean references. This design enables end-to-end learning of artifact suppression while preserving critical physiological dynamics.}
    \label{FIG:1}
\end{figure*}

Figure~\ref{FIG:1} shows the DAR pipeline. Raw EEG signals mixed with MRI-induced artifacts create the corrupted input. The encoder extracts artifact-reduced latent features used by the decoder to rebuild the clean EEG signal. This setup allows DAR to effectively remove artifacts while keeping the physiological quality of the signal, supporting further neurophysiological and clinical analyses.

\section{Materials and Methods}

To thoroughly assess how well the DAR works, we used a carefully chosen simultaneous EEG-fMRI dataset. We divided the experimental process into distinct phases: preprocessing, modeling, and evaluation.

\subsection{Dataset Description}
\label{subsec: Dataset}

This study uses the publicly available neuroimaging tools \& research collaboratory (NTRC) carbon wire loop (CWL) Eyes Open – Eyes Closed" EEG/fMRI dataset~\cite{van2016carbon}. It includes both raw and MR-corrected EEG recordings collected during simultaneous EEG-fMRI sessions. The dataset features data from eight healthy subjects recorded with a 64-channel MR-compatible EEG system (Brain Products GmbH, Gilching, Germany) in Siemens 3T MIRI scanners (TIM Trio and Verio models, Siemens Healthineer, Erlangen, Germany).

During acquisition, EEG signals were affected by various artifacts. Particularly, GA came from rapid gradient switching during echo-planar imaging (EPI). BCG artifacts resulted from cardiac pulsation and head motion within the static magnetic field. Motion-related artifacts, including those caused by helium pump vibrations, were also captured using integrated carbon wire loop motion detection channels.

The EEG recordings cover multiple experimental conditions: (1) outside the scanner (artifact-free reference), (2) inside the MRI scanner without scanning, and (3) inside the scanner during scanning with helium pump ON and OFF. These variations allow for a thorough analysis of artifact profiles. For this study, only EEG segments recorded during scanning were used to closely mimic realistic EEG-fMRI acquisition scenarios during model training and evaluation.

One subject had an inconsistent channel configuration (37 instead of 38 usable channels) and was excluded from the analysis. As a result, data from seven subjects were used in all experiments. The dataset includes manually corrected EEG references, allowing for supervised training and precise evaluation of artifact removal performance.

\subsection{EEG Data Preprocessing and Segmentation}
\label{sec: Data Preprocessing}

All EEG recordings were resampled to 250 Hz to standardize temporal resolution across subjects and reduce computational load while preserving the spectral features of typical EEG rhythms. To improve signal quality, recordings were bandpass filtered between 1 Hz and 40 Hz using a zero-phase finite impulse response (FIR) filter. The 1 Hz lower cutoff removed slow drifts and baseline wander, while the 40 Hz upper cutoff reduced high-frequency noise and residual gradient artifacts from MRI acquisition.

After filtering, continuous EEG data were segmented into overlapping windows of 2 seconds with a stride of 1 second. This segmentation provided enough temporal context for learning artifact patterns, increased training sample diversity, and captured temporal variability across sessions. Each segment contained 38 EEG channels and 500 time points (2 s $\times$ 250 Hz), forming a consistent input shape for the model.

To reduce amplitude-related biases and stabilize training, each segment was normalized to the range [-1, 1] based on its maximum absolute amplitude. This scaling ensured a uniform input distribution, allowing the model to focus on artifact-related features instead of amplitude variations.

Only segments with both artifact-contaminated signals and their manually corrected counterparts were kept to support supervised learning. This selective pairing allowed the model to learn a valid mapping from noisy to clean EEG signals, which is essential for effective artifact suppression. After preprocessing and filtering, a total of 2,163 segments were included in the final dataset for training and evaluation.

\begin{figure}[htbp]
    \centering
    \includegraphics[width=0.45\textwidth, height=0.45\textheight]{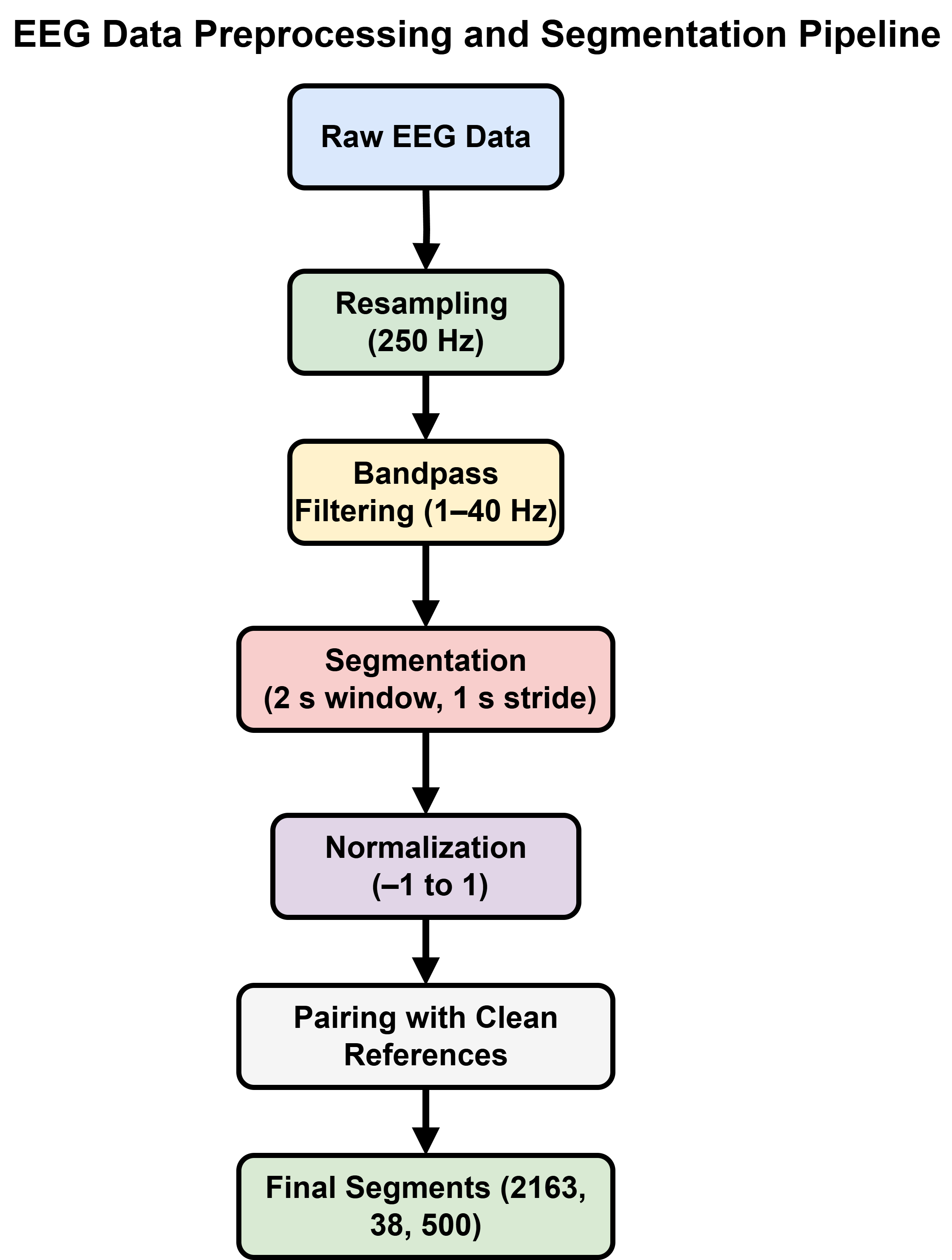}
    \caption{Schematic diagram of the EEG data preprocessing and segmentation pipeline.}
    \label{FIG:2}
\end{figure}

Although the dataset contains 2,163 segments, each provides rich spatiotemporal information (38 channels, 500 time points). The use of overlapping segmentation and robust normalization decreases the risk of underrepresentation and supports generalizable artifact correction. A detailed overview of the preprocessing pipeline is shown in Figure~\ref{FIG:2}.

\subsection{Model Architecture}
\label{subsec: Model Architecture}

The DAR serves as a 1D convolutional autoencoder focused on removing prominent GA and BCG artifacts from EEG signals captured during concurrent fMRI scans while retaining delicate neural oscillations intact in the process. The EEG segments are portrayed as ($C\times T$) matrices with $C=38$ channels and $T=500$ time points spanned across $2$ seconds at a frequency of $250$ Hz, allowing for artifact correction through the utilization of channel-specific characteristics.

The encoder module of the DAR compresses each input channel into a latent representation that highlights artifact structure while keeping relevant neural information. It has two convolutional layers:
\begin{itemize}
    \item The first layer uses 128 filters with a kernel size of 5 and padding of 2 to preserve the temporal length $T$. It applies ReLU activation followed by batch normalization for reliable and nonlinear feature learning.
    \item The second layer reduces the feature dimension to 64 channels, using the same kernel size and padding, and also follows with ReLU activation and batch normalization.
\end{itemize}
This leads to an intermediate feature map of shape $64\times T$, capturing both localized artifact features and broader temporal dependencies needed for effective denoising.

The decoder module of the DRA reflects the encoder to reconstruct the clean EEG signal from the latent representation:
\begin{itemize}
    \item The first decoder layer maintains 64 channels, using the same kernel size and padding, together with ReLU activation and batch normalization.
    \item The second layer expands the representation to 128 channels, preparing for the final transformation while applying ReLU activation and batch normalization.
    \item A final output convolutional layer reduces the 128-channel signal to a single-channel output for each EEG input, using a Tanh activation to keep output values within the normalized range $[-1, 1]$.
\end{itemize}

The reconstructed output keeps the original $C\times T$ shape, ensuring it is suitable for neurophysiological and clinical analyses. To avoid over-smoothing and maintain subtle temporal oscillations critical for cognitive and clinical interpretation, no dropout layers or weight decay are included. Instead, batch normalization is used to stabilize training. The model relies solely on localized convolutional operations, making it efficient while preserving important oscillatory content.

\begin{figure*}[htbp]
    \centering
    \includegraphics[width=\textwidth]{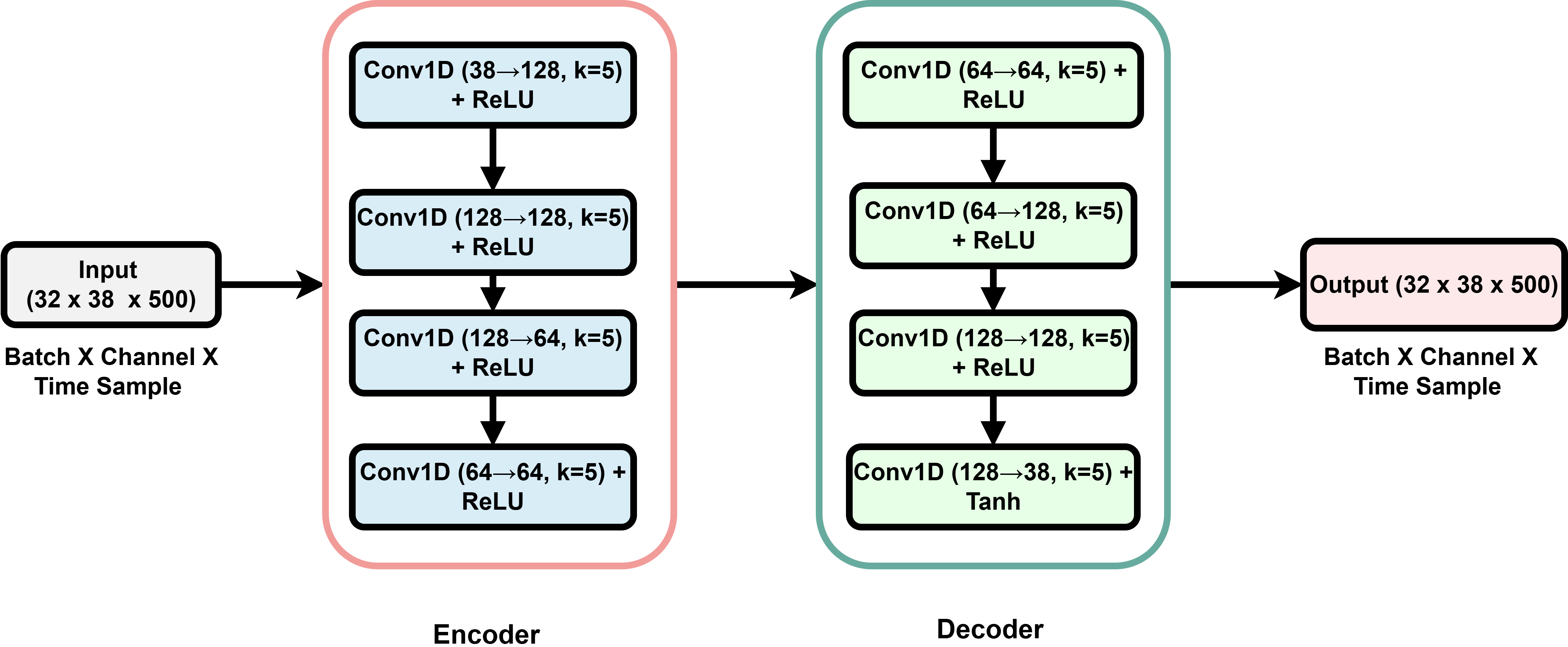}
    \caption{Detailed architecture of the proposed DAR model. The network consists of an encoder and decoder with symmetric 1D convolutional layers. The encoder uses filters of sizes 128 and 64 (kernel size = 5, padding = 2) with ReLU activations to extract latent features. The decoder reconstructs artifact-free EEG using matching layers and a final \texttt{tanh} activation to constrain output amplitudes. The DAR performs channel-wise nonlinear mapping to suppress gradient and BCG artifacts while preserving physiological signals.}
    \label{FIG:3}
\end{figure*}

Figure~\ref{FIG:3} shows the DAR architecture. It illustrates the channel-wise 1D convolutional design, including the symmetry of the encoder and decoder, the layer-wise feature expansion and reduction, the activation functions (ReLU for intermediate layers, Tanh for output), and the flow from artifact-contaminated input to denoised output. This design allows DAR to achieve strong artifact suppression without sacrificing the integrity of the underlying EEG signals, making it ideal for both neuroscience research and clinical applications.

\subsection{Training Procedure}
\label{subsec: Training Procedure}

DAR is trained with a supervised learning approach, using paired EEG segments containing artifact-contaminated inputs and their corresponding clean references. This allows the model to learn a direct mapping from noisy to artifact-free signals while keeping essential neurophysiological information.

The model is optimized with the L1 loss function (mean absolute error), chosen for its ability to handle outliers and its tendency to produce smoother and more realistic reconstructions, critical for maintaining subtle neural oscillations in EEG data~\cite{redina2025analyzing}. The loss function is defined as:

\begin{equation}
\mathcal{L}(\Theta) = \frac{1}{N} \sum_{i=1}^{N} \left\| x_i - \hat{x}_i \right\|_1
\end{equation}

where $x_i$ represents the clean reference EEG segment, $\hat{x}_i$ is the corresponding reconstructed output from the model, and $N$ is the number of training samples. The mapping function $f_\Theta$ is defined by the network weights $\Theta$.

Optimization is done using the Adam algorithm, chosen for its adaptive learning rate and fast convergence in high-dimensional parameter spaces. The learning rate is set to $1 \times 10^{-3}$, with exponential decay rates $\beta_1 = 0.9$ and $\beta_2 = 0.999$, and an epsilon value of $1 \times 10^{-8}$ for numerical stability. No weight decay is applied to maintain subtle signal dynamics.

Training uses a batch size of 32 for up to 200 epochs. Early stopping is applied with a patience of 20 epochs and a minimum validation loss improvement threshold $\Delta=1 \times 10^{-5}$ to prevent overfitting and improve convergence efficiency.

All experiments are carried out in PyTorch 2.0.1 with CUDA 12.1 for GPU support. A fixed random seed of 42 ensures consistent results across training runs. To maintain the authentic features of artifact-contaminated EEG, no data augmentation is used. This prevents introducing synthetic noise patterns, allowing the model to learn from naturally occurring MRI-induced artifacts. A summary of key training configurations and hyperparameters is provided in Table 1 for clarity and reproducibility.

\begin{figure}[htbp]
    \centering
    \includegraphics[width=0.40\textwidth, height=0.65\textheight]{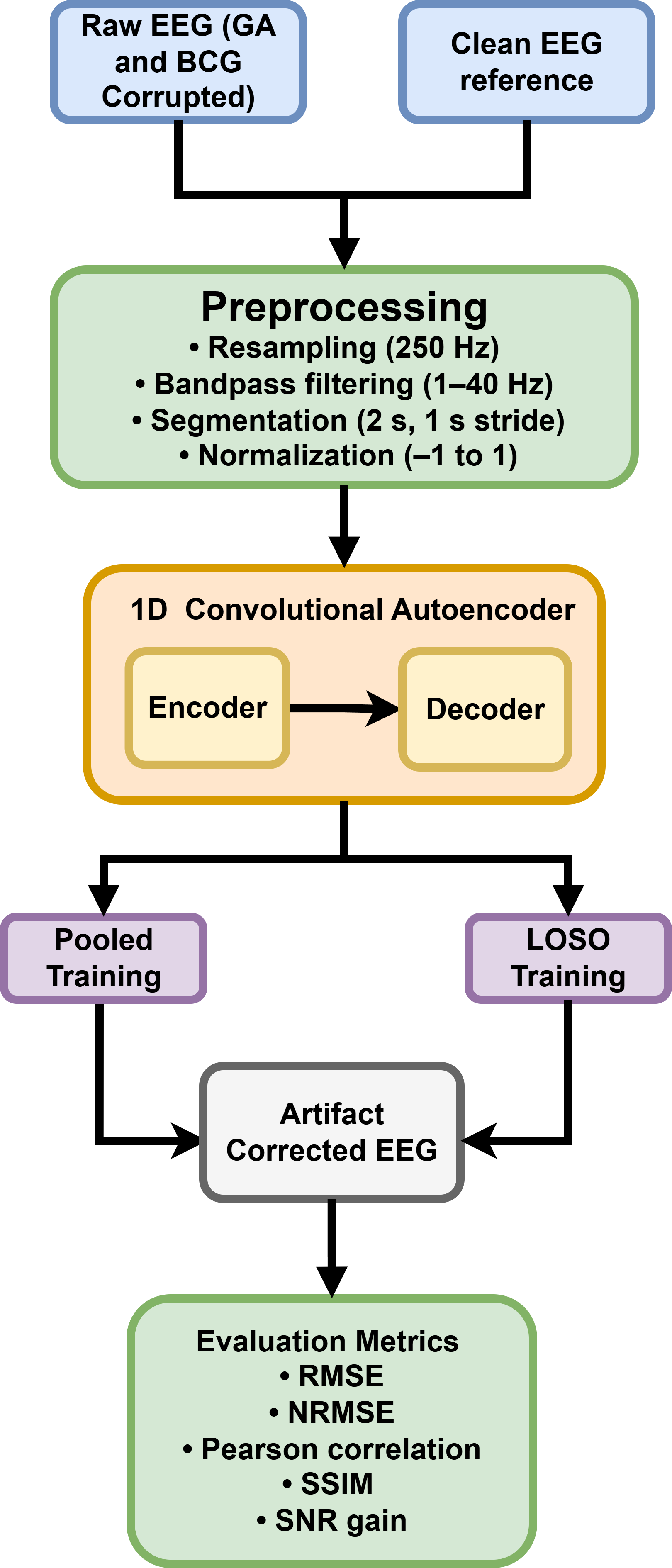}
    \caption{Overview of the proposed DAR-based EEG artifacts removal pipeline. Raw artifact-corrupted and clean EEG signals undergo resampling, bandpass filtering, segmentation, and normalization. Preprocessed inputs are passed through DAR, a 1D convolutional autoencoder for gradient and BCG artifact removal. Training is performed using pooled and leave-one-subject-out (LOSO) strategies. Denoised outputs are evaluated using RMSE, NRMSE, Pearson correlation, SSIM, and SNR gain.}
    \label{FIG:4}
\end{figure}

Table~\ref{tbl1} summarizes the key hyperparameters and settings used during model training. Figure~\ref{FIG:4} provides a visual overview of the workflow for training and evaluation.

\begin{table}[h]
\centering
\caption{DAR Training Hyperparameters and Configurations}
\label{tbl1}
\begin{tabular}{ll}
\toprule
\textbf{Parameter} & \textbf{Value / Setting} \\
\midrule
Loss function         & L1 loss (mean absolute error) \\
Optimizer             & Adam \\
Learning rate         & $1 \times 10^{-3}$ \\
Betas                 & $(0.9,\ 0.999)$ \\
Epsilon               & $1 \times 10^{-8}$ \\
Weight decay          & None \\
Batch size            & 32 \\
Maximum epochs        & 200 \\
Early stopping        & Patience = 20,\ $\Delta = 1 \times 10^{-5}$ \\
Framework             & PyTorch 2.0.1 (CUDA 12.1) \\
Random seed           & 42 \\
Data splits           & Pooled (80/20), LOSO cross-validation \\
Data augmentation     & None \\
\bottomrule
\end{tabular}
\end{table}

Figure ~\ref{FIG:4} illustrates the overall methodology of the work in a constructive manner.

Two data splitting strategies are used to evaluate both within-subject and cross-subject generalization:
\begin{itemize}
    \item \textbf{Pooled Training:} Data from all subjects are combined and randomly split into 80\% for training and 20\% for validation. This simulates a setting where the model learns generalizable artifact patterns across subjects.
    \item \textbf{Leave-One-Subject-Out (LOSO) Cross-Validation:} Each subject is iteratively held out as the test set, while the model is trained on the remaining subjects. This rigorous approach assesses the model's ability to generalize to unseen individuals~\cite{kunjan2021necessity}.
\end{itemize}

\subsection{Evaluation Metrics}
\label{subsec: Evaluation Matrices}

The performance of DAR was evaluated using six metrics: root mean squared error (RMSE), normalized RMSE (NRMSE), mean absolute error (MAE), Pearson correlation coefficient, structural similarity index (SSIM), signal-to-noise ratio (SNR) gain, and cosine similarity. These metrics provide a thorough assessment of the model's effectiveness in reducing artifacts while preserving the underlying EEG signal.

\textbf{Root Mean Squared Error (RMSE):}

\begin{equation}
\text{RMSE} = \sqrt{ \frac{1}{N} \sum_{i=1}^{N} (x_i - \hat{x}_i)^2 } \tag{3} \\
\end{equation}

\textbf{Normalized Root Mean Square Error (NRMSE):}
\begin{equation}
\text{NRMSE} = \frac{\text{RMSE}}{\max(x) - \min(x)} \tag{4} 
\end{equation}

\textbf{Mean Absolute Error (MAE):} 
\begin{equation}
\text{MAE} = \frac{1}{N} \sum_{i=1}^{N} |x_i - \hat{x}_i| \tag{5}
\end{equation}

\textbf{Pearson Correlation Coefficient:}
\begin{equation}
r_{xy} = \frac{ \sum (x_i - \bar{x})(y_i - \bar{y}) }{ \sqrt{ \sum (x_i - \bar{x})^2 } \sqrt{ \sum (y_i - \bar{y})^2 } } \tag{6}
\end{equation}

\textbf{Structural Similarity Index Measure:}
\begin{equation}
\text{SSIM}(x, \hat{x}) = \frac{ (2\mu_x \mu_{\hat{x}} + C_1)(2\sigma_{x\hat{x}} + C_2) }{ (\mu_x^2 + \mu_{\hat{x}}^2 + C_1)(\sigma_x^2 + \sigma_{\hat{x}}^2 + C_2) } \tag{7}
\end{equation}

\textbf{Signal-to-Noise Ratio:}
\begin{equation}
\text{SNR} = 10 \log_{10} \left( \frac{ \sum x_i^2 }{ \sum (x_i - \hat{x}_i)^2 } \right) \tag{8}
\end{equation}

\textbf{Cosine Similarity}
\begin{equation}
\text{Cosine Similarity} = \frac{ \sum_{i=1}^{N} x_i \hat{x}_i }{ \sqrt{ \sum_{i=1}^{N} x_i^2 } \sqrt{ \sum_{i=1}^{N} \hat{x}_i^2 } } \tag{9}
\end{equation}

\subsection{Quntitative Comparison}
\label{subsec: Quntitative Comparison}
To evaluate DAR's artifacts removal performance in comparison with available deep learning models, a statistical significance analysis was conducted using paired $t$-tests on the validation dataset. The method was evaluated on BiLSTM, ResNet1D, and U-Net1D using two core metrics: RMSE and Pearson correlation coefficient.

\subsection{Ablation Study}
\label{subsec: Ablation Study}
In order to evaluate the impact of key architectural components in DAR, an ablation study was conducted by systematically modifying its structure. For consistency, all ablation variants were trained for 50 epochs without aggressive fine-tuning or early stopping, ensuring a controlled comparison unaffected by training duration or hyperparameter sensitivity.

\subsection{Saliency-Based Interpretation}
\label{subsec: Saliency-Based Interpretation}

To better understand DAR’s internal decision-making process and its ability to focus on relevant EEG regions, a saliency-based interpretability analysis was performed. Gradients of the model output with respect to the input were computed during pooled training, allowing us to visualize the temporal and spatial contributions of individual EEG channels and time points to the final denoised output.

\subsection{Statistical and Interpretive Analysis}
\label{subsec: Statistical Analysis}

Due to evaluate how DAR performed in comparison with the established deep learning baseline models, including BiLSTM, ResNet1D, and U-Net1D, we conducted a statistical significance analysis using paired $t$-tests. We emphasize two key metrics: root mean squared error (RMSE) and Pearson correlation coefficient.

To interpret the model's internal behavior further, we performed a saliency-based analysis. This involved calculating input gradients with respect to model outputs during pooled training. We aimed to determine which temporal and spatial features had the most influence on artefact removal.

\subsection{Subject-Wise Generalization Study}
\label{subsec: Generalization Study}

To evaluate generalization across individuals with varying artifact characteristics, a LOSO cross-validation protocol was employed. In each iteration, DAR was trained on EEG recordings from six subjects and tested on the held-out subject. This process was repeated so that each subject served once as the test set.

\section{Results}
\label{sec: Results}

This section presents the comprehensive evaluation of the DAR. The results include quantitative comparisons with baseline denoising methods, statistical significance testing, ablation studies, and interpretability analyses. DAR demonstrates considerable artifact suppression while preserving essential EEG features.

\subsection{Quantitative Denoising Performance}
\label{subsec: Quantitative Performance}

DAR was evaluated on the validation set using several standard metrics to quantify denoising performance. Table~\ref{tbl2} summarizes the results across all key metrics, including RMSE, NRMSE, Pearson correlation coefficient, SSIM, MAE, cosine similarity, and SNR gain. We also analyzed the power spectral density (PSD) of the denoised EEG signals to assess spectral fidelity and artifact suppression in the frequency domain. This analysis is visualized in supplementary Figure~\ref{Supp:1} 

\begin{table}[h!]
\centering
\caption{Artifact Removal Performance of the Proposed DAR Model on the Validation Set.}
\label{tbl2}
\begin{tabular}{ll}
\toprule
\textbf{Metric} & \textbf{Value (Mean $\pm$ SD)} \\
\midrule
RMSE $\downarrow$ & $0.0218 \pm 0.0152$ \\
NRMSE $\downarrow$ & $0.0139 \pm 0.0085$ \\
MAE $\downarrow$ & $0.00540$ \\
Pearson Corr. $\uparrow$ & $0.8799 \pm 0.2238$ \\
Cosine Similarity $\uparrow$ & $0.9363$ \\
SSIM $\uparrow$ & $0.8885 \pm 0.0913$ \\
SNR Gain (dB) $\uparrow$ & $+14.63$ \\
\bottomrule
\end{tabular}
\end{table}

The DAR achieved an average RMSE of $0.0218 \pm 0.0152$, NRMSE of $0.0139 \pm 0.0085$, and SSIM of $0.8885 \pm 0.0913$. The Pearson correlation between the denoised and reference EEG signals was $0.8799 \pm 0.2238$, indicating high temporal fidelity. Additionally, the MAE was $0.00540$, and the cosine similarity reached $0.9363$, confirming waveform preservation. The model also achieved an average SNR gain of $+14.63$~dB, demonstrating substantial improvement in signal clarity.

\subsection{Comparison with Traditional Baseline Methods}
\label{subsec: Comparison with Traditional}

To evaluate the effectiveness of DAR, we compared its performance with widely used traditional EEG denoising techniques, including PCA, AAS, ICA, and OBS. Table~\ref{tbl3} presents the comparative results on the validation set.

DAR generally performed better than other traditional methods (~\ref{tbl3}) across most of the evaluation metrics. Notably, ICA, commonly regarded as one of the most effective unsupervised methods, achieved an RMSE of $0.0585 \pm 0.0198$ and an SSIM of $0.6346 \pm 0.1236$, both showed comparatively lower performance than DAR. Similarly, PCA, AAS, and OBS exhibited higher reconstruction errors and poorer structural similarity, highlighting the limitations of these methods in capturing complex artifact patterns. These findings demonstrate the enhanced artifacts removal capability and structural preservation of the deep learning-based DAR model.

\begin{table*}[t]
\centering
\caption{Performance Comparison with Traditional Baselines on the Validation Set}
\label{tbl3}
\begin{tabular}{lcccc}
\toprule
\textbf{Method} & \textbf{RMSE} & \textbf{NRMSE} & \textbf{Correlation} & \textbf{SSIM} \\
\midrule
DAR & $0.0218 \pm 0.0152$ & $0.0139 \pm 0.0085$ & $0.8799 \pm 0.2238$ & $0.8885 \pm 0.0913$ \\
PCA     & $0.1013 \pm 0.0298$ & $0.0668 \pm 0.0188$ & $0.1726 \pm 0.2512$ & $0.4898 \pm 0.1641$ \\
AAS     & $0.0990 \pm 0.0304$ & $0.0654 \pm 0.0194$ & $0.2010 \pm 0.2500$ & $0.5579 \pm 0.1553$ \\
ICA     & $0.0585 \pm 0.0198$ & $0.0377 \pm 0.0091$ & $0.5100 \pm 0.0820$ & $0.6346 \pm 0.1236$ \\
OBS     & $0.0991 \pm 0.0262$ & $0.0655 \pm 0.0162$ & $0.1984 \pm 0.2293$ & $0.5565 \pm 0.1482$ \\
\bottomrule
\end{tabular}
\end{table*}

\subsection{Comparison with Deep Learning-Based Approaches}
\label{subsec: Comparison with DL}

DAR not only performs better than traditional artifact removal methods but also outperforms several recent deep learning-based EEG denoising architectures, including BiLSTM, ResNet1D, Transformer, Temporal Convolutional Network (TCN), 1D U-Net, MSDCNet, WaveNet Denoiser, and Attention U-Net 1D. 

Table~\ref{tbl4} reports the performance of all models using MAE, cosine similarity, and SNR gain. DAR achieved the best overall performance with the lowest MAE ($0.0054$), highest cosine similarity ($0.9363$), and largest SNR gain ($+14.63$~dB). Among other models, 1D U-Net showed relatively competitive performance, achieving a cosine similarity of $0.9052$ and an SNR gain of $+12.67$~dB. However, all compared architectures fell short of DAR's overall denoising accuracy and signal fidelity.

\begin{table}[h!]
\centering
\caption{Performance Comparison with Deep Learning Models on the Validation Set}
\label{tbl4}
\begin{tabular}{lccc}
\toprule
\textbf{Model} & \makecell{\textbf{MAE}} & \makecell{\textbf{Cosine}\\\textbf{Similarity}} & \makecell{\textbf{SNR Gain}\\\textbf{(dB)}} \\
\midrule
DAR                 & $0.0054$ & $0.9363$ & $+14.63$ \\
BiLSTM              & $0.0121$ & $0.8325$ & $+9.24$ \\
ResNet1D            & $0.0102$ & $0.8596$ & $+10.32$ \\
Transformer         & $0.0087$ & $0.8831$ & $+11.56$ \\
TCN                   & $0.0109$ & $0.8490$ & $+9.98$ \\
1D U-Net            & $0.0061$ & $0.9052$ & $+12.67$ \\
MSDCNet             & $0.0093$ & $0.8744$ & $+10.94$ \\
WaveNet Denoiser    & $0.0075$ & $0.8905$ & $+11.82$ \\
Attention U-Net 1D  & $0.0068$ & $0.8979$ & $+12.13$ \\
\bottomrule
\end{tabular}
\end{table}

\subsection{Statistical Significance Analysis}
\label{subsec: Statistical Significance Analysis}

Table~\ref{tbl5} summarizes the complete results of the statistical analysis. For RMSE, DAR demonstrated significantly lower reconstruction errors across all comparisons. For instance, the $t$-statistic against BiLSTM was $-6.73$ with a $p$-value $< 0.0001$, and Cohen’s $d = 1.30$, indicating a large effect size. Similar patterns were observed when compared with ResNet1D and U-Net1D, with $t$-statistics less than $-6.0$ and $d>1.2$.

For correlation, DAR demonstrated comparatively higher similarity scores than the baseline DL models. The $t$-statistic was $6.77$ for BiLSTM, and Cohen’s $d = 1.31$, with all comparisons yielding $p<0.0001$ and $d>1.3$, confirming the model’s ability to preserve EEG waveform structures post-denoising.

\begin{table}[h!]
\centering
\caption{Statistical significance analysis comparing DAR with BiLSTM, ResNet1D, and U-Net1D across RMSE and correlation. All $p$-values are $< 0.0001$ with large effect sizes (Cohen’s $d>1$).}
\label{tbl5}
\begin{tabular}{lcccc}
\toprule
\textbf{Metric} & \makecell{\textbf{Compared}\\\textbf{Method}} & \makecell{\textbf{$t$-}\\\textbf{statistic}} & \makecell{\textbf{$p$-}\\\textbf{value}} & \makecell{\textbf{Cohen’s}\\\textbf{$d$}} \\
\midrule
\multirow{5}{*}{RMSE}
& BiLSTM   & $-6.73$  & $< 0.0001$ & $1.30$ \\
& ResNet1D & $-6.18$  & $< 0.0001$ & $1.20$ \\
& U-Net1D  & $-6.25$  & $< 0.0001$ & $1.21$ \\
\midrule
\multirow{3}{*}{Correlation}
& BiLSTM   & $6.77$   & $< 0.0001$ & $1.31$ \\
& ResNet1D & $6.71$   & $< 0.0001$ & $1.30$ \\
& U-Net1D  & $6.97$   & $< 0.0001$ & $1.34$ \\
\bottomrule
\end{tabular}
\end{table}

\subsection{Insights from Ablation Study}
\label{subsec:Ablation}

\begin{figure*}[htbp]
    \centering
    \includegraphics[width= 0.7\textwidth]{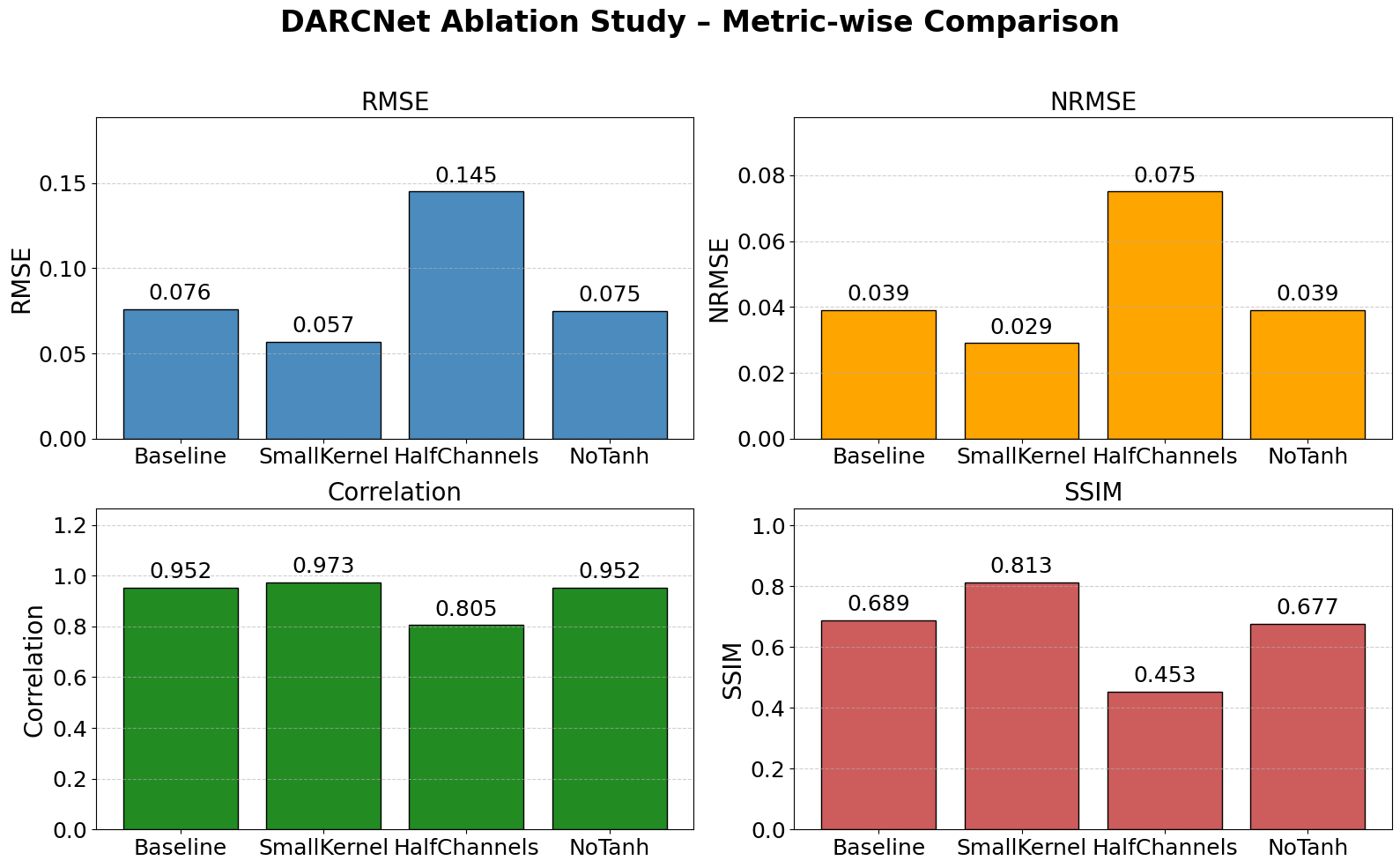}
    \caption{Metric-wise comparison from the DAR ablation study. Performance metrics (RMSE, NRMSE, Pearson correlation, SSIM) were evaluated for the baseline DAR model and three variants. HalfChannels significantly degrades all metrics. SmallKernel improves RMSE and SSIM. NoTanh slightly lowers SSIM with negligible impact on RMSE or correlation. Kernel size, channel count, and output activation are crucial for EEG denoising.}
    \label{Supp:2}
\end{figure*}

Table~\ref{tbl6} presents ablation study results for three architectural modifications: (1) reducing the convolutional kernel size (\textit{SmallKernel}), (2) halving the number of feature channels (\textit{HalfChannels}), and (3) removing the final Tanh activation function (\textit{NoTanh}). The baseline corresponds to the original DAR configuration under the same limited training conditions. Supplementary Figure~\ref{Supp:2} provides visual interpretations corresponding to these configurations.

Reducing the number of channels substantially degraded performance, with RMSE rising to 0.1451 and SSIM dropping to 0.4528, emphasizing the importance of sufficient channel capacity for modeling complex EEG dynamics. In contrast, using smaller kernels improved SSIM to 0.8134 and reduced RMSE to 0.0568, suggesting that finer local feature extraction benefits artifact suppression. Removing the Tanh activation led to slightly higher RMSE and a decrease in SSIM to 0.6771, indicating its role in output amplitude control and reconstruction stability.

It is important to note that the results in Table~\ref{tbl6} are not directly comparable to fully optimized models presented in Tables~\ref{tbl2}–\ref{tbl5}, as the ablation experiments focus solely on architectural effects under controlled settings.

\begin{table}[h!]
\centering
\caption{Ablation study results for DAR architectural variants. Metrics include RMSE, NRMSE, Pearson correlation, and SSIM. Each variant isolates a key component’s effect on performance.}
\label{tbl6}
\begin{tabular}{lcccc}
\toprule
\textbf{Variant} & \textbf{RMSE} & \textbf{NRMSE} & \textbf{Correlation} & \textbf{SSIM} \\
\midrule
Baseline       & 0.0757 & 0.0391 & 0.9516 & 0.6895 \\
SmallKernel    & 0.0568 & 0.0293 & 0.9734 & 0.8134 \\
HalfChannels   & 0.1451 & 0.0749 & 0.8052 & 0.4528 \\
NoTanh         & 0.0748 & 0.0386 & 0.9523 & 0.6771 \\
\bottomrule
\end{tabular}
\end{table}

\subsection{Interpretability Analysis}
\label{subsec: Interpretibility Analysis}

\begin{figure*}[htbp]
    \centering
    \includegraphics[width=\textwidth]{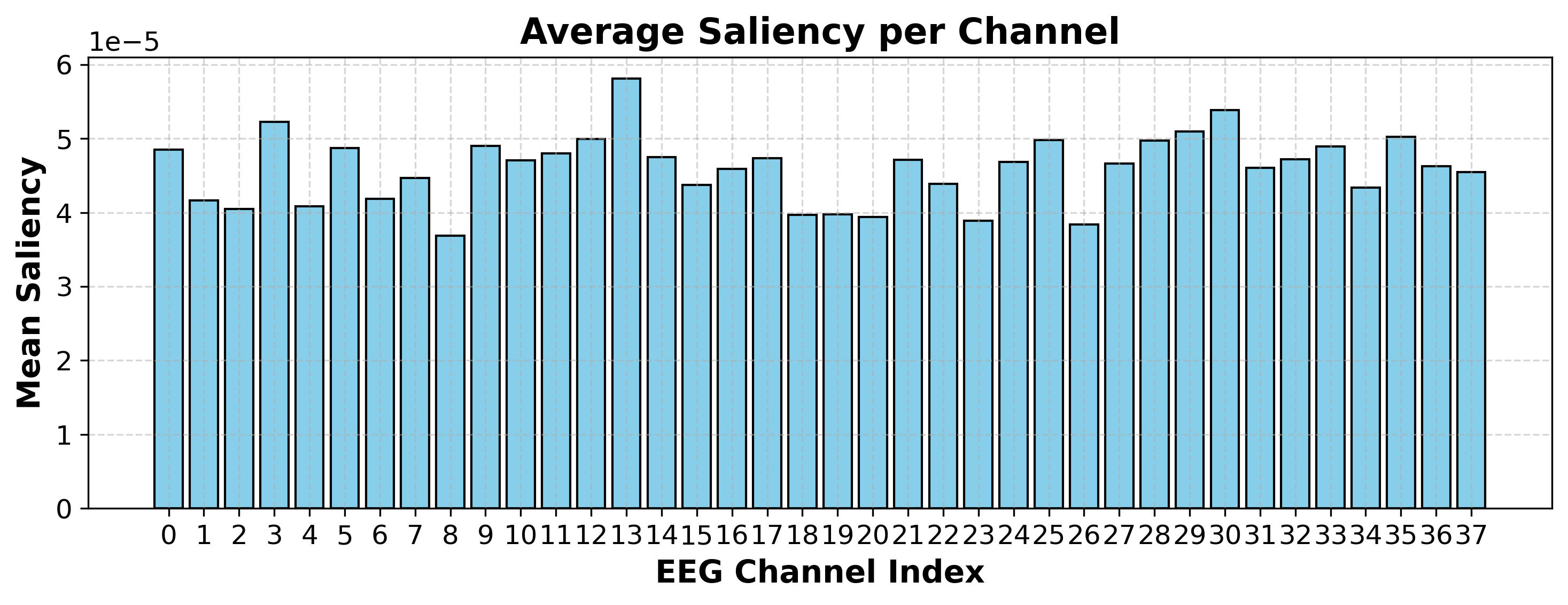}
    \caption{An interpretability assessment of the DAR model took place using saliency-based visualizations. The team calculated the average importance of each channel across all validation samples. This approach highlighted the channels that had the biggest impact on the model's denoising decisions.}
    \label{FIG:5}
\end{figure*}

\begin{figure}[htbp]
    \centering
    \includegraphics[width= 0.5\textwidth, height=0.22\textheight]{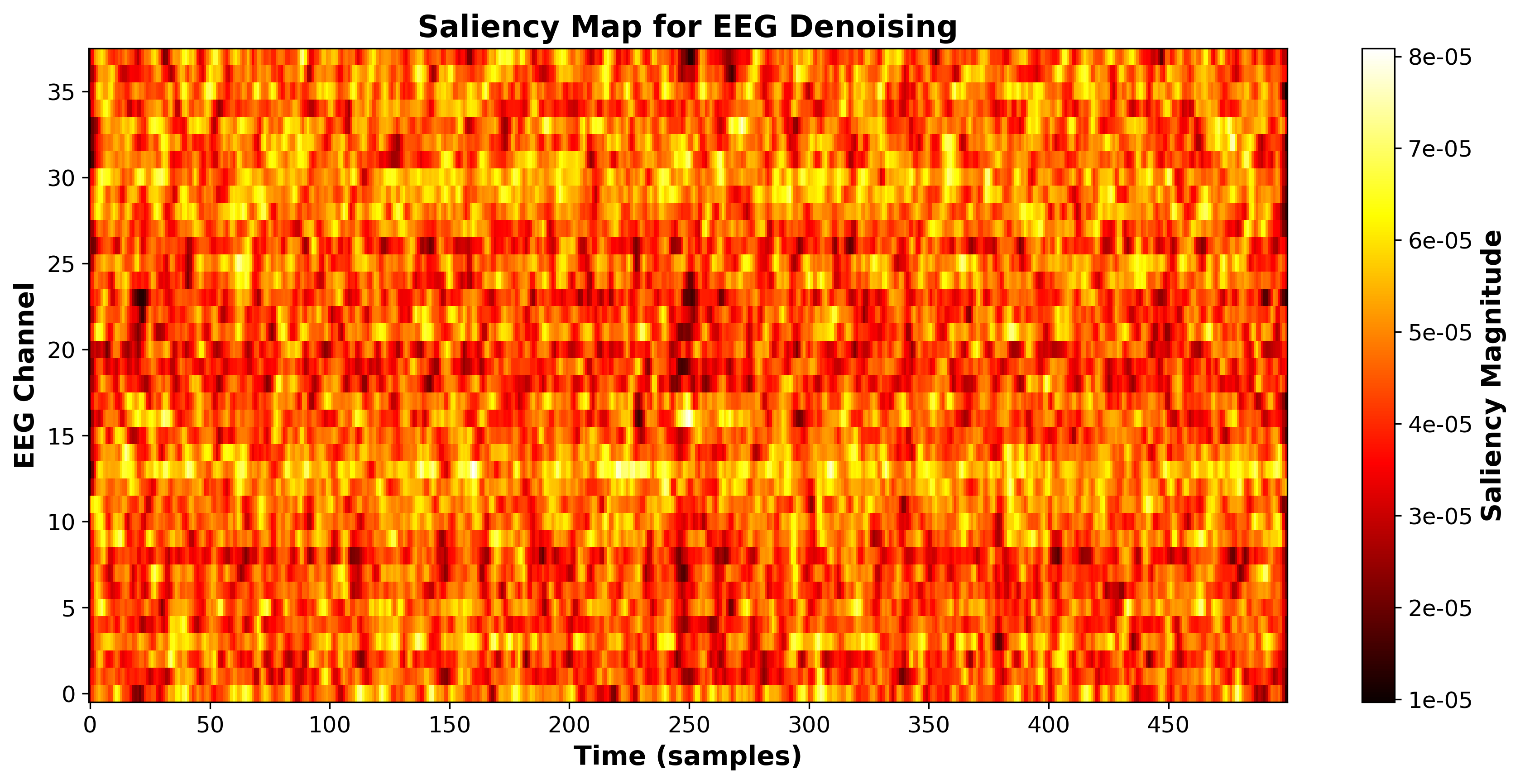}
    \caption{Interpretability analysis of DAR using saliency-based visualization. Channel–time saliency heatmap for a representative sample, highlighting localized temporal contributions across channels.}
    \label{Supp:3}
\end{figure}

Figure~\ref{FIG:5} shows the average saliency across 38 EEG channels. It highlights how each channel affects the model's decision-making. Peaks in saliency indicate that some channels are more relevant than others. The highest contribution is seen near channel 13. Moreover, supplementary Figure~\ref{Supp:3} features a saliency heatmap across channels and time for a representative sample

\begin{figure*}[htbp]
    \centering
   \includegraphics[width=1.2\textwidth]{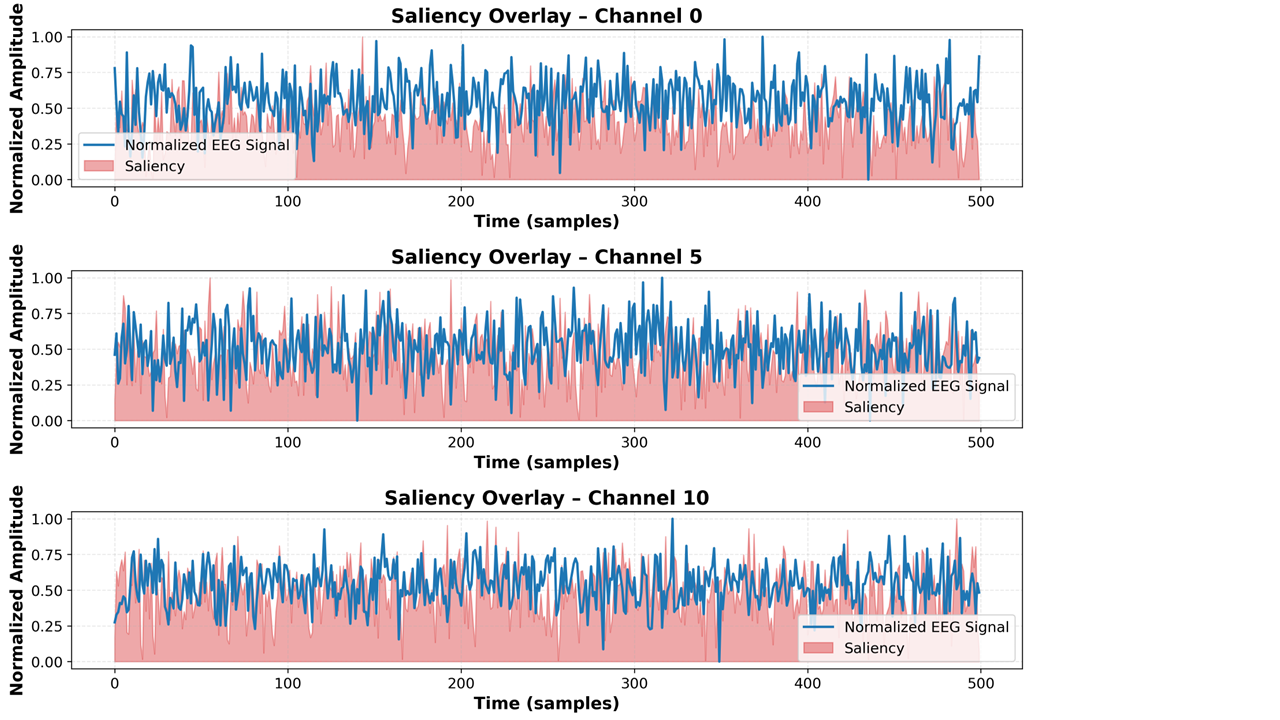}
    \caption{Saliency-based interpretability results across three EEG channels: Channel 0 (top), Channel 5 (middle), and Channel 10 (bottom). Each subplot shows the normalized EEG signal (blue line) and the corresponding saliency magnitude (red shaded area), revealing the temporal importance learned by the model for artifact correction.}
    \label{FIG:6}
\end{figure*}

Figure~\ref{FIG:6}  illustrates saliency overlays for three representative EEG channels (Channel 0 – top, Channel 5 – middle, and Channel 10 – bottom), where the raw normalized EEG signals are superimposed with the corresponding saliency magnitudes.

The blue line indicates the normalized EEG signal, while the red shaded region reflects the saliency values, highlighting regions of high relevance as determined by the model's gradients. In each channel, we observe that the model consistently attributes higher importance (denser saliency) to specific temporal segments, particularly around signal fluctuations. This behavior confirms that DAR attends substantially to portions of the signal where artifacts are likely to occur or have been corrected. These insights validate the model's temporal sensitivity and its ability to focus on critical EEG dynamics for artifact removal.

\subsection{Qualitative Analysis}
\label{subsec: Qualitative Analysis}

\begin{figure*}[htbp]
    \centering
    \includegraphics[width=1.6\textwidth]{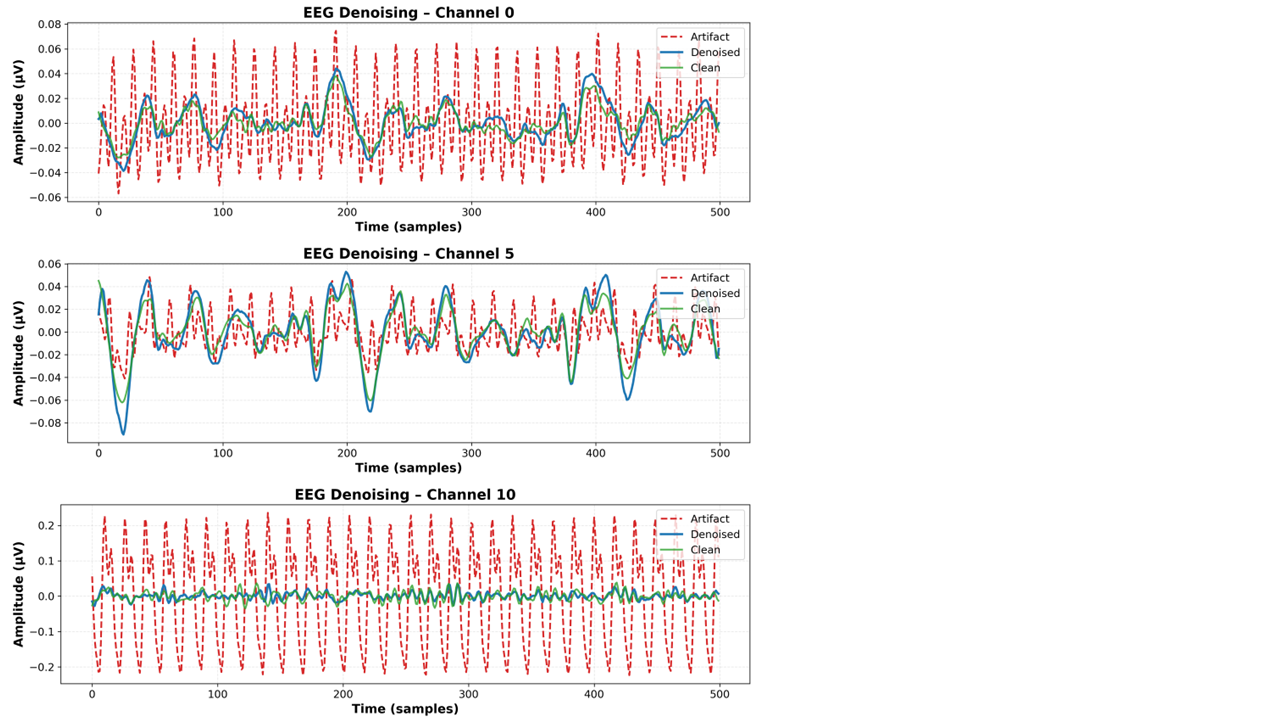}
    \caption{Channel-wise EEG denoising examples using DAR. Overlays of artifact-corrupted (blue dashed), ground truth clean (green), and DAR-denoised (orange) signals are shown for three representative channels: Channel 0 (top), Channel 5 (middle), and Channel 10 (bottom). DAR effectively suppresses high-amplitude artifacts while preserving key temporal and morphological features of the original EEG signals.}
    \label{FIG:7}
\end{figure*}

In addition to quantitative evaluation, a qualitative analysis was conducted to visually assess DAR's artifacts removal performance and its ability to preserve underlying EEG waveform integrity. Figure~\ref{FIG:7} presents EEG waveforms from three selected channels (0, 5, and 10), including artifact-corrupted inputs, clean references, and DAR-denoised outputs.

In all cases, DAR effectively suppresses high-amplitude artifacts while closely tracking the timing and shape of the clean EEG signals. These examples demonstrate that the model retains key oscillatory patterns necessary for neurophysiological interpretation, rather than simply smoothing the signal. A broader comparison across additional channels is provided in Supplementary Figure~\ref{Supp:4}.

\begin{figure}[htbp]
    \centering
    \includegraphics[width=0.45\textwidth]{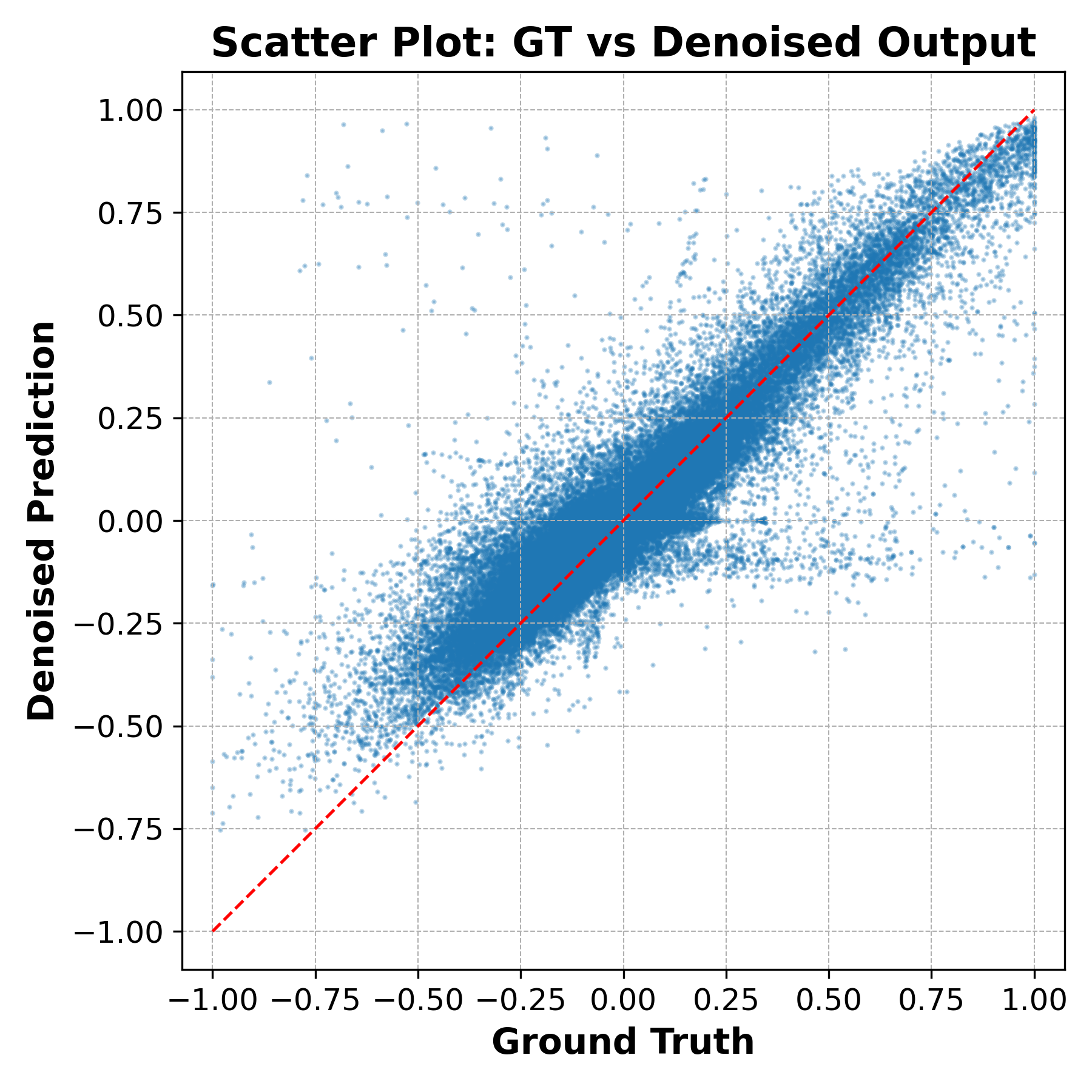}
    \caption{Scatter plot comparing ground truth (GT) amplitudes and DAR-denoised predictions. Each point represents a sample across all channels and time points in a representative validation segment. The close clustering around the identity line (red dashed) indicates strong agreement and confirms the high fidelity of the denoised outputs.}
    \label{FIG:8}
\end{figure}

To further confirm signal reconstruction fidelity, Figure~\ref{FIG:8} presents a scatter plot comparing clean and predicted amplitudes across all samples and time points from a representative validation segment. The tight clustering along the identity line reflects strong amplitude and phase preservation, confirming DAR’s high reconstruction accuracy and minimal bias.

These qualitative findings reinforce the quantitative results, providing visual evidence of DAR's ability to effectively remove EEG artifacts while preserving meaningful signal content.

\subsection{Subject-Wise Generalization (LOSO Cross-Validation)}
\label{subsec:LOSO}

Table~\ref{tbl7} summarizes performance across all subjects using RMSE, Pearson correlation, and SSIM. DAR consistently performed well across individuals, achieving an average RMSE of $0.0635 \pm 0.0110$, average correlation of $0.4428 \pm 0.2063$, and average SSIM of $0.6658 \pm 0.0880$. These results highlight DAR's robustness to inter-subject variability in both artifact intensity and EEG morphology.

\begin{table*}[h!]
\centering
\caption{Per-subject performance metrics in LOSO cross-validation. DAR consistently achieves low RMSE, moderate-to-high correlation, and strong SSIM across subjects.}
\label{tbl7}
\begin{tabular}{lccc}
\toprule
\textbf{Subject} & \textbf{RMSE} & \textbf{Correlation} & \textbf{SSIM} \\
\midrule
0 & 0.0841 & 0.3898 & 0.5257 \\
1 & 0.0578 & 0.8316 & 0.6445 \\
2 & 0.0686 & 0.2455 & 0.7442 \\
3 & 0.0721 & 0.1355 & 0.5929 \\
4 & 0.0579 & 0.4729 & 0.7482 \\
5 & 0.0509 & 0.5056 & 0.6218 \\
6 & 0.0532 & 0.5189 & 0.7837 \\
\midrule
\textbf{Mean $\pm$ SD} & \textbf{0.0635 $\pm$ 0.0110} & \textbf{0.4428 $\pm$ 0.2063} & \textbf{0.6658 $\pm$ 0.0880} \\
\bottomrule
\end{tabular}
\end{table*}

To statistically validate the LOSO results, a Shapiro–Wilk test~\cite{razali2011power} confirmed normality of the RMSE, correlation, and SSIM distributions ($p>0.43$), permitting the use of parametric analyses. Additionally, 95\% confidence intervals were estimated via bootstrap resampling~\cite{hesterberg2015teachers} ($N=1000$), yielding:  
\textbf{RMSE:} [0.056, 0.072],  
\textbf{Correlation:} [0.300, 0.595],  
\textbf{SSIM:} [0.602, 0.730].

\begin{figure}[htbp]
    \centering
    \includegraphics[width=0.45\textwidth]{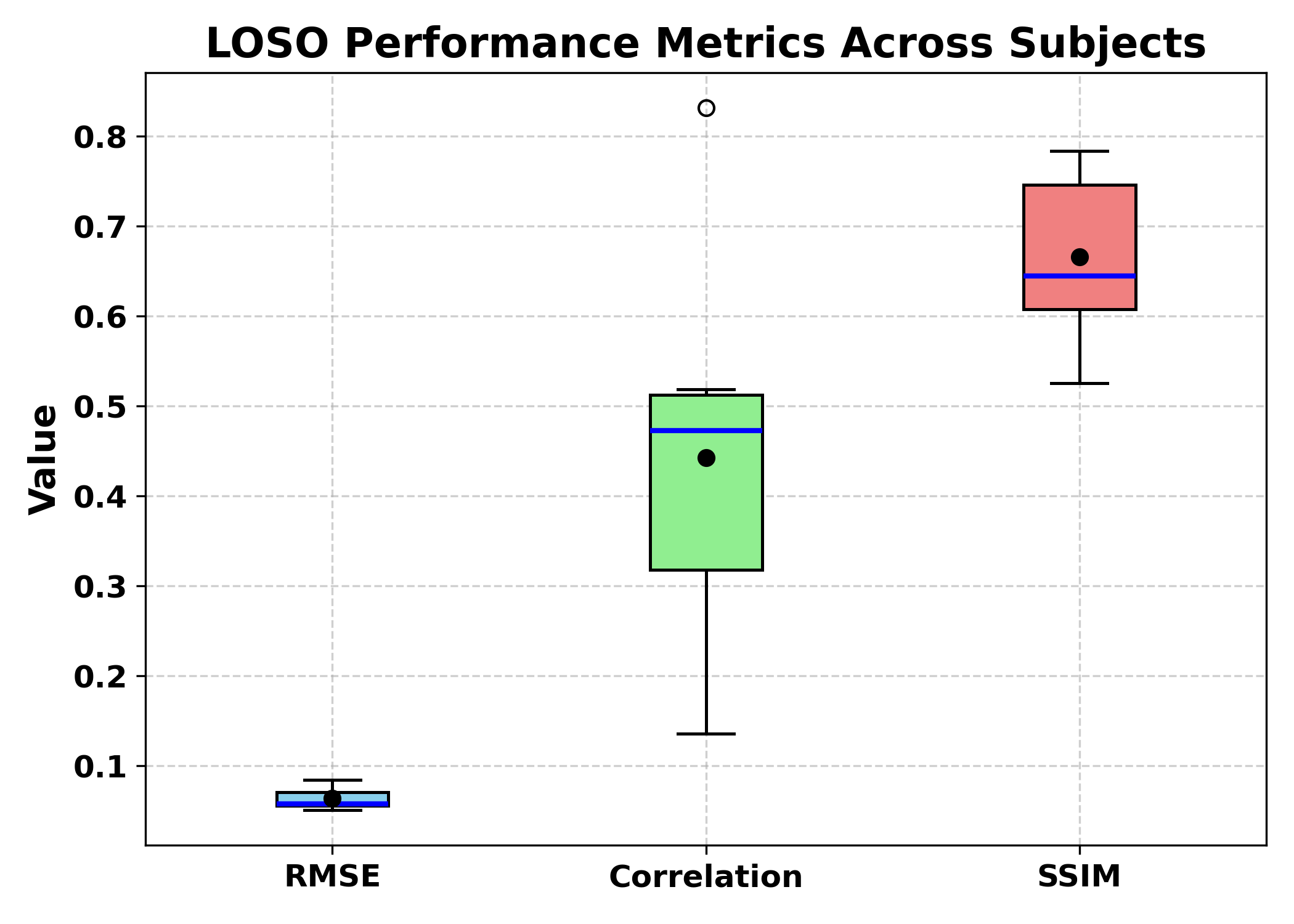}
    \caption{Boxplots of LOSO performance metrics across subjects. Distribution of RMSE, Pearson correlation coefficient, and SSIM scores achieved by DAR during leave-one-subject-out cross-validation. Each box represents inter-subject variability, with black dots indicating mean values and whiskers showing data range. The results highlight DAR’s consistent performance across unseen individuals, maintaining low RMSE and high correlation and SSIM scores despite subject-level differences.}
    \label{FIG:9}
\end{figure}

Figure~\ref{FIG:9} further illustrates the distribution of LOSO performance metrics across subjects using boxplots for RMSE, correlation, and SSIM. The variability is modest, indicating consistent performance across the dataset.

\section{Discussion}
\label{sec: Discussion}

This study presented DAR, a new one-dimensional convolutional autoencoder that suppresses serious artifacts found in simultaneous EEG-fMRI acquisitions while maintaining important neurophysiological dynamics. Through detailed quantitative and qualitative analyses, DAR showed better denoising performance than traditional correction methods and recent deep learning approaches.

In pooled validation analyses, DAR achieved a low RMSE of $0.0218 \pm 0.0152$ and a high SSIM of $0.8885 \pm 0.0913$, confirming its ability to reconstruct clean EEG signals with minimal distortion. The model preserved temporal structures, supported by a Pearson correlation of $0.8799 \pm 0.2238$ and a cosine similarity of $0.9363$. Additionally, the observed SNR gain of $+14.63$ dB indicated a significant improvement in signal clarity without losing important brain activity patterns.

Compared to traditional methods like ICA, AAS, OBS, and PCA \cite{gonccalves2007artifact, jafarifarmand2013artifacts, kher2016adaptive, correa2007artifact, chawla2011pca, turnip2014removal, ter2013reduction}, DAR consistently produced lower reconstruction errors and better structural similarity. Although ICA is widely used for artifact removal, it relies heavily on manual component selection and linear assumptions, which can lead to losing neural information or incomplete artifact removal. Template-based methods (like AAS and OBS) and PCA also lack the flexibility to capture the nonlinear and subject-specific features of artifacts in EEG-fMRI recordings.

In comparison, DAR’s data-driven, nonlinear design effectively models complex artifact patterns, such as those from gradient switching and ballistocardiographic (BCG) effects, without needing handcrafted features or external references. When compared with recent deep learning models like BiLSTM, ResNet1D, and 1D U-Net, DAR showed better performance across all metrics. Statistical significance analysis confirmed these results, with all $p$-values $<0.0001$ and Cohen’s $d>1.2$, indicating large effect sizes and affirming the robustness and practical relevance of these improvements \cite{mcintosh2020ballistocardiogram, yang2018automatic, lin2022ballistocardiogram, duffy2020gradient, mayeli2021automated, gao2022eeg}.

Ablation studies revealed the impact of key architectural components. Reducing channel capacity significantly lowered performance, highlighting the need for sufficient feature extraction capability. Smaller kernel sizes improved local pattern learning, which enhanced both RMSE and SSIM. The final \texttt{tanh} activation helped control output amplitude and improve structural fidelity, supporting its inclusion in the design.

To improve understanding of deep learning models, we added a saliency-based analysis. Gradient-based saliency maps showed that DAR focuses on artifact-heavy areas without negatively affecting physiological signal components. This clarity in model behavior builds confidence for clinical and neuroscientific adoption.

Generalization was tested using the LOSO cross-validation approach. DAR maintained considerable denoising performance across all subjects, with an average RMSE of $0.0635 \pm 0.0110$, a correlation of $0.4428 \pm 0.2063$, and an SSIM of $0.6658 \pm 0.0880$. Some variability in correlation values may come from subject-specific differences in artifact intensity and distribution, suggesting future improvements such as adaptive fine-tuning or transfer learning methods.

This study focused on short EEG segments (2 seconds), which may limit analyses in long-duration recordings. Future work will look into extending DAR to longer sequences. While the CWL dataset provides a solid benchmark, it has a limited sample size (seven subjects after exclusions). Broader validation on more varied datasets, especially in clinical populations, is essential to evaluate generalizability. Adding real-time denoising capability and testing in future EEG-fMRI studies will further establish DAR’s value in real-world settings.

DAR represents a promising advancement in EEG-fMRI artifact correction. Its potential denoising accuracy, preservation of neural signal structure, and added interpretability features make it a promising option for both research and clinical use. Future efforts will aim to improve its adaptability, refine its interpretability features, and validate its performance across different neuroimaging conditions.

\section{Conclusion}
\label{sec: Conclusion}

This study deployed DAR, a one-dimensional convolutional autoencoder designed to remove artifacts in simultaneous EEG-fMRI recordings. By using channel-wise nonlinear mappings and data-driven learning, DAR effectively reduces gradient and BCG artifacts while preserving important neurophysiological information.

Though it is a preliminary work based on a small dataset, quantitative results showed that DAR considerably outperformed traditional methods like ICA, PCA, AAS, and OBS, along with several top deep learning models. It achieved lower reconstruction errors, higher structural similarity, and better SNR gains. These improvements were statistically significant and backed by large effect sizes, showing both strength and practical importance.

Importantly, saliency-based interpretability analysis revealed that the model focuses on artifact-heavy regions. Additionally, LOSO cross-validation confirmed that DAR has the potential to generalize across individuals, emphasizing its usefulness in clinical and research settings.

In summary, DAR offers a promising framework for correcting artifacts in EEG-fMRI data. Future work will aim to expand its validation to a larger dataset and more varied groups, enable real-time use, and improve interpretability for clinical integration.

%\appendix
%\section{My Appendix}
%Appendix sections are coded under \verb+\appendix+.

\printcredits

\section*{Declaration of competing interest}
The authors affirm that they have no financial conflicts of interest or personal affiliations that could have potentially influenced the research findings or interpretations presented in this study.

\section*{Data availability}
The dataset is available at \url{https://www.nitrc.org/projects/cwleegfmri_data/} and is published in the study by \cite{van2016carbon}.

\section*{Acknowledgment}
The authors gratefully acknowledge the use of Grammarly for language refinement. All authors reviewed and approved the final manuscript.

\section*{Supplementary Material}

Additional visualizations and comparisons are included in the Supplementary Material to support and expand on the findings presented in the main text.

\begin{figure*}[htbp]
    \centering
    \includegraphics[width= 0.7\textwidth]{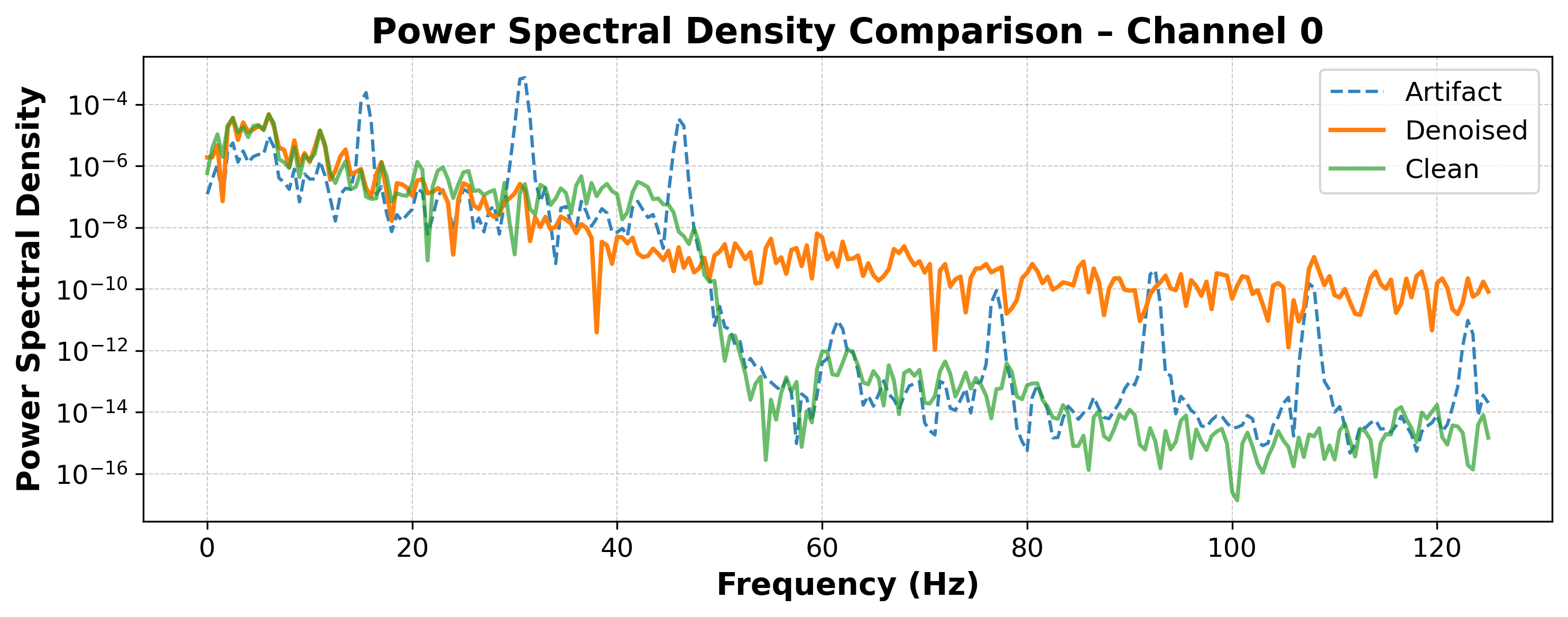}
    \caption{Power spectral density (PSD) comparison for Channel 0. The artifact-contaminated signal (blue dashed) shows prominent high-frequency peaks, while the DAR-denoised signal (orange) exhibits a spectrum closely matching the clean reference signal (green), indicating effective artifact suppression and spectral preservation.}
    \label{Supp:1}
\end{figure*}

\begin{figure*}[htbp]
    \centering
    \includegraphics[width=0.9\textwidth, height=0.9\textheight]{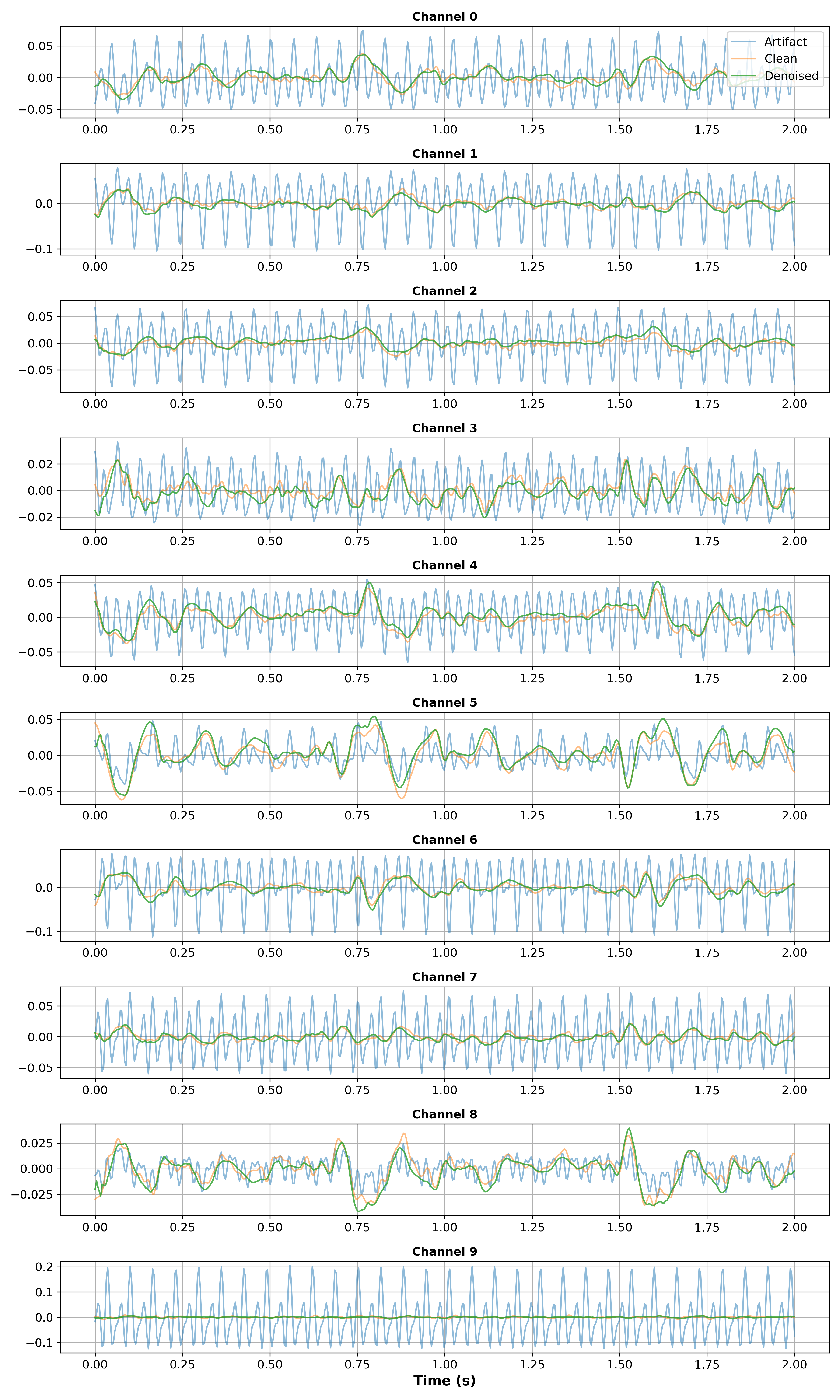}
    \caption{Channel-wise waveform comparison of artifact-contaminated EEG (blue), DAR-denoised signals (green), and clean reference signals (orange) across ten representative channels (Channels 0–9). The DAR outputs closely follow the clean references, effectively suppressing large-amplitude artifacts while preserving fine-grained oscillatory patterns and temporal dynamics critical for neurophysiological interpretation.}
    \label{Supp:4}
\end{figure*}

\FloatBarrier  % Ensures this figure stays in the results section

%\bibliographystyle{cas-model2-names}
\bibliographystyle{elsarticle-num}
%\bibliographystyle{elsarticle-harv}
%\bibliographystyle{apalike}

\bibliography{cas-refs}

\begin{thebibliography}{10}
\expandafter\ifx\csname url\endcsname\relax
  \def\url#1{\texttt{#1}}\fi
\expandafter\ifx\csname urlprefix\endcsname\relax\def\urlprefix{URL }\fi
\expandafter\ifx\csname href\endcsname\relax
  \def\href#1#2{#2} \def\path#1{#1}\fi

\bibitem{uriguen2015eeg}
J.~A. Urig{\"u}en, B.~Garcia-Zapirain, Eeg artifact removal—state-of-the-art
  and guidelines, Journal of neural engineering 12~(3) (2015) 031001.

\bibitem{roy2019deep}
Y.~Roy, H.~Banville, I.~Albuquerque, A.~Gramfort, T.~H. Falk, J.~Faubert, Deep
  learning-based electroencephalography analysis: a systematic review, Journal
  of neural engineering 16~(5) (2019) 051001.

\bibitem{freyer2009ultrahigh}
F.~Freyer, R.~Becker, K.~Anami, G.~Curio, A.~Villringer, P.~Ritter,
  Ultrahigh-frequency eeg during fmri: pushing the limits of imaging-artifact
  correction, Neuroimage 48~(1) (2009) 94--108.

\bibitem{logothetis2008we}
N.~K. Logothetis, What we can do and what we cannot do with fmri, Nature
  453~(7197) (2008) 869--878.

\bibitem{blinowska2021practical}
K.~J. Blinowska, J.~{\.Z}ygierewicz, Practical biomedical signal analysis using
  MATLAB{\textregistered}, cRc Press, 2021.

\bibitem{scrivener2021simultaneous}
C.~L. Scrivener, When is simultaneous recording necessary? a guide for
  researchers considering combined eeg-fmri, Frontiers in Neuroscience 15
  (2021) 636424.

\bibitem{mulert2013simultaneous}
C.~Mulert, Simultaneous eeg and fmri: towards the characterization of structure
  and dynamics of brain networks, Dialogues in clinical neuroscience 15~(3)
  (2013) 381--386.

\bibitem{wu2016real}
X.~Wu, T.~Wu, Z.~Zhan, L.~Yao, X.~Wen, A real-time method to reduce
  ballistocardiogram artifacts from eeg during fmri based on optimal basis sets
  (obs), Computer methods and programs in biomedicine 127 (2016) 114--125.

\bibitem{yan2009understanding}
W.~X. Yan, K.~J. Mullinger, M.~J. Brookes, R.~Bowtell, Understanding gradient
  artefacts in simultaneous eeg/fmri, Neuroimage 46~(2) (2009) 459--471.

\bibitem{mullinger2013best}
K.~J. Mullinger, P.~Castellone, R.~Bowtell, Best current practice for obtaining
  high quality eeg data during simultaneous fmri, Journal of visualized
  experiments: JoVE~(76) (2013) 50283.

\bibitem{gonccalves2007artifact}
S.~Gon{\c{c}}alves, P.~Pouwels, J.~Kuijer, R.~Heethaar, J.~De~Munck, Artifact
  removal in co-registered eeg/fmri by selective average subtraction, Clinical
  Neurophysiology 118~(11) (2007) 2437--2450.

\bibitem{jafarifarmand2013artifacts}
A.~Jafarifarmand, M.~A. Badamchizadeh, Artifacts removal in eeg signal using a
  new neural network enhanced adaptive filter, Neurocomputing 103 (2013)
  222--231.

\bibitem{kher2016adaptive}
R.~Kher, R.~Gandhi, Adaptive filtering based artifact removal from
  electroencephalogram (eeg) signals, in: 2016 International Conference on
  Communication and Signal Processing (ICCSP), IEEE, 2016, pp. 0561--0564.

\bibitem{correa2007artifact}
A.~G. Correa, E.~Laciar, H.~Pati{\~n}o, M.~Valentinuzzi, Artifact removal from
  eeg signals using adaptive filters in cascade, in: Journal of Physics:
  Conference Series, Vol.~90, IOP Publishing, 2007, p. 012081.

\bibitem{chawla2011pca}
M.~Chawla, Pca and ica processing methods for removal of artifacts and noise in
  electrocardiograms: A survey and comparison, Applied Soft Computing 11~(2)
  (2011) 2216--2226.

\bibitem{turnip2014removal}
A.~Turnip, E.~Junaidi, Removal artifacts from eeg signal using independent
  component analysis and principal component analysis, in: 2014 2nd
  International Conference on Technology, Informatics, Management, Engineering
  \& Environment, IEEE, 2014, pp. 296--302.

\bibitem{ter2013reduction}
E.~M. ter Braack, B.~de~Jonge, M.~J. Van~Putten, Reduction of tms induced
  artifacts in eeg using principal component analysis, IEEE transactions on
  neural systems and rehabilitation engineering 21~(3) (2013) 376--382.

\bibitem{mcintosh2020ballistocardiogram}
J.~R. McIntosh, J.~Yao, L.~Hong, J.~Faller, P.~Sajda, Ballistocardiogram
  artifact reduction in simultaneous eeg-fmri using deep learning, IEEE
  Transactions on Biomedical Engineering 68~(1) (2020) 78--89.

\bibitem{yang2018automatic}
B.~Yang, K.~Duan, C.~Fan, C.~Hu, J.~Wang, Automatic ocular artifacts removal in
  eeg using deep learning, Biomedical Signal Processing and Control 43 (2018)
  148--158.

\bibitem{lin2022ballistocardiogram}
G.~Lin, J.~Zhang, Y.~Liu, T.~Gao, W.~Kong, X.~Lei, T.~Qiu, Ballistocardiogram
  artifact removal in simultaneous eeg-fmri using generative adversarial
  network, Journal of Neuroscience Methods 371 (2022) 109498.

\bibitem{duffy2020gradient}
B.~A. Duffy, A.~W. Toga, H.~Kim, Gradient artifact correction for simultaneous
  eeg-fmri using denoising autoencoders, in: 2020 IEEE 17th International
  Symposium on Biomedical Imaging (ISBI), IEEE, 2020, pp. 1--4.

\bibitem{mayeli2021automated}
A.~Mayeli, O.~Al~Zoubi, K.~Henry, C.~K. Wong, E.~J. White, Q.~Luo, V.~Zotev,
  H.~Refai, J.~Bodurka, T.~. Investigators, et~al., Automated pipeline for eeg
  artifact reduction (appear) recorded during fmri, Journal of neural
  engineering 18~(4) (2021) 0460b4.

\bibitem{gao2022eeg}
T.~Gao, D.~Chen, Y.~Tang, Z.~Ming, X.~Li, Eeg reconstruction with a dual-scale
  cnn-lstm model for deep artifact removal, IEEE Journal of Biomedical and
  Health Informatics 27~(3) (2022) 1283--1294.

\bibitem{chuang2022ic}
C.-H. Chuang, K.-Y. Chang, C.-S. Huang, T.-P. Jung, Ic-u-net: a u-net-based
  denoising autoencoder using mixtures of independent components for automatic
  eeg artifact removal, NeuroImage 263 (2022) 119586.

\bibitem{samek2017explainable}
W.~Samek, T.~Wiegand, K.-R. M{\"u}ller, Explainable artificial intelligence:
  Understanding, visualizing and interpreting deep learning models, arXiv
  preprint arXiv:1708.08296 (2017).

\bibitem{tjoa2020survey}
E.~Tjoa, C.~Guan, A survey on explainable artificial intelligence (xai): Toward
  medical xai, IEEE transactions on neural networks and learning systems
  32~(11) (2020) 4793--4813.

\bibitem{allen2000method}
P.~J. Allen, O.~Josephs, R.~Turner, A method for removing imaging artifact from
  continuous eeg recorded during functional mri, Neuroimage 12~(2) (2000)
  230--239.

\bibitem{niazy2005removal}
R.~K. Niazy, C.~F. Beckmann, G.~D. Iannetti, J.~M. Brady, S.~M. Smith, Removal
  of fmri environment artifacts from eeg data using optimal basis sets,
  Neuroimage 28~(3) (2005) 720--737.

\bibitem{srivastava2005ica}
G.~Srivastava, S.~Crottaz-Herbette, K.~Lau, G.~H. Glover, V.~Menon, Ica-based
  procedures for removing ballistocardiogram artifacts from eeg data acquired
  in the mri scanner, Neuroimage 24~(1) (2005) 50--60.

\bibitem{debener2008properties}
S.~Debener, K.~J. Mullinger, R.~K. Niazy, R.~W. Bowtell, Properties of the
  ballistocardiogram artefact as revealed by eeg recordings at 1.5, 3 and 7 t
  static magnetic field strength, International Journal of Psychophysiology
  67~(3) (2008) 189--199.

\bibitem{bank2023autoencoders}
D.~Bank, N.~Koenigstein, R.~Giryes, Autoencoders, Machine learning for data
  science handbook: data mining and knowledge discovery handbook (2023)
  353--374.

\bibitem{van2016carbon}
J.~N. van~der Meer, A.~Pampel, E.~J. Van~Someren, J.~R. Ramautar, Y.~D. van~der
  Werf, G.~Gomez-Herrero, J.~Lepsien, L.~Hellrung, H.~Hinrichs, H.~E.
  M{\"o}ller, et~al., Carbon-wire loop based artifact correction outperforms
  post-processing eeg/fmri corrections—a validation of a real-time
  simultaneous eeg/fmri correction method, Neuroimage 125 (2016) 880--894.

\bibitem{redina2025analyzing}
R.~Redina, J.~Hejc, M.~Filipenska, Z.~Starek, Analyzing the performance of
  biomedical time-series segmentation with electrophysiology data, Scientific
  Reports 15~(1) (2025) 11776.

\bibitem{kunjan2021necessity}
S.~Kunjan, T.~S. Grummett, K.~J. Pope, D.~M. Powers, S.~P. Fitzgibbon,
  T.~Bastiampillai, M.~Battersby, T.~W. Lewis, The necessity of leave one
  subject out (loso) cross validation for eeg disease diagnosis, in:
  International conference on brain informatics, Springer, 2021, pp. 558--567.

\bibitem{razali2011power}
N.~M. Razali, Y.~B. Wah, et~al., Power comparisons of shapiro-wilk,
  kolmogorov-smirnov, lilliefors and anderson-darling tests, Journal of
  statistical modeling and analytics 2~(1) (2011) 21--33.

\bibitem{hesterberg2015teachers}
T.~C. Hesterberg, What teachers should know about the bootstrap: Resampling in
  the undergraduate statistics curriculum, The american statistician 69~(4)
  (2015) 371--386.

\end{thebibliography}


%Author biography without author photo.
%Author biography. Author biography. Author biography.
%Author biography. Author biography. Author biography.
%Author biography. Author biography. Author biography.
%Author biography. Author biography. Author biography.
%Author biography. Author biography. Author biography.
%Author biography. Author biography. Author biography.
%Author biography. Author biography. Author biography.
%Author biography. Author biography. Author biography.
%Author biography. Author biography. Author biography.

%\bio{figs/cas-pic1}
%Author biography with author photo.
%Author biography. Author biography. Author biography.
%%%Author biography. Author biography. Author biography.
%%Author biography. Author biography. Author biography.
%Author biography. Author biography. Author biography.
%Author biography. Author biography. Author biography.
%Author biography. Author biography. Author biography.
%Author biography. Author biography. Author biography.
%Author biography. Author biography. Author biography.
%Author biography. Author biography. Author biography.

%\bio{figs/cas-pic1}
%Author biography with author photo.
%Author biography. Author biography. Author biography.
%Author biography. Author biography. Author biography.
%Author biography. Author biography. Author biography.
%Author biography. Author biography. Author biography.

\end{document}